\documentclass[aps,prb,twocolumn,superscriptaddress,floatfix]{revtex4-2}
\usepackage{amsmath,amssymb,amsfonts}
\usepackage{ragged2e}
\usepackage{graphicx}
\usepackage{xcolor}
\usepackage{hyperref}
\usepackage{braket}
\usepackage{physics}
\usepackage{mathtools}
\usepackage{bbm}
\usepackage{dsfont}
\usepackage{booktabs}

\newcommand{\rOT}{\mathcal{R}_{\mathrm{OT}}}
\usepackage[justification=justified]{subcaption}
\usepackage[justification=justified]{caption}
\usepackage{soul} 
\newcommand{\Ustar}{\mathcal{U}^*}
\newcommand{\Lhop}{\mathcal{L}_{\mathrm{hop}}}
\newcommand{\Lint}{\mathcal{L}_{\mathrm{int}}}

\begin{document}

\title{A Geometric Theory of Fermion-to-Qubit Encodings}

\author{Lakshya Nagpal}
\email{lakshyan@imsc.res.in}
\affiliation{Institute of Mathematical Sciences, CIT Campus, Chennai 600113, India}
\affiliation{QCAR Group, The Institute of Mathematical Sciences (IMSc), Chennai 600113, India}
\affiliation{Pecslab Research}

\author{Nishith Reen}
\affiliation{Institute of Mathematical Sciences, CIT Campus, Chennai 600113, India}
\affiliation{QCAR Group, The Institute of Mathematical Sciences (IMSc), Chennai 600113, India}

\author{S. R. Hassan}
\affiliation{Institute of Mathematical Sciences, CIT Campus, Chennai 600113, India}
\affiliation{QCAR Group, The Institute of Mathematical Sciences (IMSc), Chennai 600113, India}
\affiliation{Homi Bhabha National Institute, Anushakti Nagar, Mumbai, Maharashtra 400094}

\date{\today}

\begin{abstract}
Exact fermion-to-qubit transformations are conventionally regarded as
algorithmic tools that translate many-body Hamiltonians into qubit
representations for quantum simulation. Here we show that they also define
intrinsic geometric representations whose structure encodes physically
meaningful information beyond spectral equivalence. We develop a geometric
framework based on weighted hypergraphs and coupling-space representations
constructed from the Bravyi--Kitaev (BK) and Xia--Bian--Kais (XBK)
encodings~\cite{Nag26}. Within the BK representation, we introduce a geometric observable
that compares the algebraic connectivities of the kinetic and interaction
hypergraphs, derive its exact analytical dependence on interaction strength,
and uncover two geometric universality classes together with an exact
spectral organization originating from the binary-tree architecture of the
encoding. The complementary XBK representation describes the evolution of
encoded Hamiltonians through probability measures in coupling space, where
optimal transport quantifies interaction-driven reorganization independently
of the spectral analysis. Applications to the Hubbard, spinless $t$--$V$,
single-impurity Anderson, and Kitaev models demonstrate that these
connectivity- and transport-based geometric descriptions consistently
capture the structural evolution of encoded quantum Hamiltonians across
distinct classes of many-body systems. Our results establish hypergraph
geometry as a new framework for understanding fermion-to-qubit encodings,
revealing that they serve not only as computational mappings but also as
geometric representations of quantum many-body Hamiltonians.
\end{abstract}

\maketitle

\section{Introduction}

Every exact representation of a physical system preserves its spectrum, but
not every representation reveals the same physical structure. Throughout the
history of physics, new mathematical representations have repeatedly changed
the way physical phenomena are understood. Phase space reformulated classical
mechanics in terms of geometry, Hilbert space transformed quantum mechanics
into a theory of linear operators, and reciprocal space revealed the
organization of crystalline solids. These representations do more than
provide equivalent mathematical descriptions; they expose hidden structure,
suggest new observables, and often reshape physical intuition. This raises a
natural question for quantum computation: can an exact fermion-to-qubit
encoding reveal physical structure beyond simply reproducing the spectrum of
the original many-body Hamiltonian?

Fermion-to-qubit transformations have become indispensable for quantum
computation~\cite{Fey99,Llo96,Pre18,NielsenChuang2010} by providing exact mappings between interacting fermionic
Hamiltonians and qubit operators. Among the most widely used are the
Jordan--Wigner, parity, and Bravyi--Kitaev (BK) transformations~\cite{Jor28,Bra00,See12}, each of
which preserves the complete many-body spectrum while offering different
trade-offs in locality and operator complexity. Consequently, these mappings
have been studied almost exclusively from an algorithmic perspective, where
their primary purpose is to prepare Hamiltonians for quantum simulation~\cite{Asp05,Per14,Cao19,Mca18}.
Their quality is therefore assessed in terms of computational resources,
including operator locality, qubit requirements, circuit depth, and gate
count~\cite{Hav17,See12,Ste19}.

This computational viewpoint leaves open a more fundamental question. Does a
fermion-to-qubit transformation merely provide an equivalent encoding of the
original Hamiltonian, or does the encoding itself possess an intrinsic
geometric organization carrying physical information? If such a geometry
exists, it should emerge naturally from the encoded Hamiltonian itself,
independent of the many-body wavefunction or the solution of the underlying
quantum problem.

In this work we demonstrate that such a geometry indeed emerges naturally.
Every Pauli string generated by a fermion-to-qubit transformation defines a
multi-qubit interaction, and the complete encoded Hamiltonian therefore
induces a weighted hypergraph whose structure evolves with the physical
parameters of the underlying model. This observation shifts the focus from
the quantum state to the encoded Hamiltonian itself. Rather than asking how
the many-body wavefunction changes with interaction strength, we investigate
how the geometry induced by the encoding reorganizes as the Hamiltonian is
varied.

The Hubbard model provides an ideal setting for exploring this idea. As the
paradigmatic model of strongly correlated electrons, it describes the
competition between kinetic delocalization and Coulomb repulsion that
underlies a broad range of interaction-driven phenomena. Conventionally,
this competition is characterized through observables derived from the
many-body wavefunction, such as correlation functions, spectral functions,
or transport properties~\cite{Whi92,Whi93,Sch11,Oru14,Tro05}. Here we show that complementary information is
already encoded in the geometry induced by the fermion-to-qubit
representation, allowing interaction-driven reorganization to be studied
without direct reference to the exponentially large Hilbert space.

To develop this framework, we combine two complementary representations of
the encoded Hamiltonian. The Bravyi--Kitaev (BK) representation~\cite{Bra00,See12} naturally
induces a hypergraph whose spectral properties characterize the connectivity
of the encoding, while the Xia--Bian--Kais (XBK) representation~\cite{Xia17,Xia21} provides a
complementary description of the evolution of the complete coupling
distribution through an exactly equivalent diagonal Ising Hamiltonian.
Together, these representations establish two complementary geometric
viewpoints: one based on spectral connectivity and the other on
coupling-space transport.

Within this unified framework we uncover several unexpected structural
properties of fermion-to-qubit encodings. We derive an exact geometric
observable whose interaction dependence follows a closed analytical form,
identify two universality classes associated with tapered and untapered BK
encodings, discover an exact spectral partition governed by the binary-tree
backbone of the BK transformation, and show that the same interaction-driven
reorganization is independently captured through optimal transport in
coupling space. Extending the analysis to the Hubbard, spinless $t$--$V$,
Anderson impurity, and Kitaev models further demonstrates that these
geometric signatures are not restricted to a particular Hamiltonian but
reflect a broader organization of correlated quantum systems.

Our results suggest that fermion-to-qubit mappings contain substantially
more physical information than required for quantum simulation alone.
Beyond preserving the spectrum, they naturally induce geometric structures
whose organization reflects the underlying many-body physics. This
establishes hypergraph geometry as a general framework for connecting quantum
encodings, spectral geometry, optimal transport, and strongly correlated
quantum matter.

\section{A Geometric Theory of Fermion-to-Qubit Encodings}
\label{sec:framework}

The central idea of this work is that every exact fermion-to-qubit encoding
naturally defines a geometric object associated with the encoded Hamiltonian.
Conventionally, an encoding is viewed as a map that preserves the spectrum of
a fermionic Hamiltonian while expressing it in terms of qubit operators.
Here we adopt a different viewpoint. We regard the encoded Hamiltonian itself
as the primary object and ask whether its operator structure possesses an
intrinsic geometry capable of revealing physical information independently of
the many-body wavefunction.

Consider a generic interacting fermionic Hamiltonian

\begin{equation}
H_f(\lambda),
\end{equation}

where $\lambda$ denotes the set of physical parameters. An exact
fermion-to-qubit transformation produces an equivalent qubit Hamiltonian

\begin{equation}
H_f(\lambda)
\longrightarrow
H_q(\lambda),
\end{equation}

while preserving the complete many-body spectrum.

Irrespective of the particular encoding, the qubit Hamiltonian admits the
general Pauli decomposition

\begin{equation}
H_q
=
\sum_{\alpha=1}^{N_P}
J_\alpha(\lambda)\,
P_\alpha,
\label{eq:general_pauli}
\end{equation}

where $P_\alpha$ denotes a tensor product of single-qubit Pauli operators,
$J_\alpha(\lambda)$ is its coupling coefficient, and $N_P$ is the total
number of Pauli strings appearing in the encoded Hamiltonian~\cite{NielsenChuang2010}.

Equation~(\ref{eq:general_pauli}) defines a natural geometric
representation. Every qubit is associated with a vertex, while every Pauli
string defines a multi-qubit interaction among the qubits on which it acts
nontrivially. Since a Pauli string may involve more than two qubits, the
appropriate mathematical object is a weighted hypergraph rather than an
ordinary graph.

We therefore associate every encoded Hamiltonian with the weighted
hypergraph

\[
\mathcal{H}=(V,E,w),
\]

defined by

\[
V=\{q_i\},
\qquad
E=\{e_\alpha\}_{\alpha=1}^{N_P},
\]

where each hyperedge $e_\alpha$ corresponds uniquely to the Pauli string
$P_\alpha$ appearing in Eq.~(\ref{eq:general_pauli}). The weight function

\[
w:E\rightarrow\mathbb{R}^{+}
\]

assigns to every hyperedge the magnitude of its corresponding coupling,

\[
w(e_\alpha)=|J_\alpha|.
\]

The mapping

\[
H_q
\longrightarrow
\mathcal{H}
\]

is exact and preserves the complete operator content of the encoded
Hamiltonian. Single-qubit operators generate single-vertex hyperedges,
two-qubit operators generate ordinary edges, and higher-order Pauli strings
naturally generate higher-order hyperedges. Consequently, every encoded
Hamiltonian defines a unique weighted hypergraph whose geometry is determined
entirely by its operator structure.

\subsection{Spectral Geometry of Encoded Hamiltonians}
\label{sec:spectral_geometry}

The weighted hypergraph introduced above provides a complete geometric
representation of the encoded Hamiltonian. To extract quantitative
information from this geometry, we require observables that characterize its
global organization. Throughout this work these observables are obtained from
the spectrum of an associated graph Laplacian.

We first map the weighted hypergraph onto an equivalent weighted graph using
the standard clique expansion, in which every hyperedge is replaced by a
complete weighted clique preserving its contribution to the overall
connectivity. The resulting weighted adjacency matrix,
$\mathbf{A}$,
defines the degree matrix

\[
D_{ii}
=
\sum_j A_{ij},
\]

from which the combinatorial graph Laplacian is constructed as

\[
L=D-A.
\]

The Laplacian spectrum,

\[
0=\lambda_1
<
\lambda_2
\le
\cdots
\le
\lambda_n,
\]

provides a hierarchy of geometric invariants associated with the encoded
Hamiltonian. Rather than representing energies, these eigenvalues quantify
the organization of the interaction network defined by the Pauli operators.

Among them, the second-smallest eigenvalue,

\[
\lambda_2(L),
\]

known as the Fiedler eigenvalue or algebraic connectivity, plays a central
role. It measures the global coherence of the network and therefore serves
as the simplest spectral descriptor of the geometry of the encoded
Hamiltonian. Changes in the Hamiltonian parameters modify the coupling
coefficients $\{J_\alpha\}$, thereby changing the hyperedge weights,
altering the Laplacian, and ultimately shifting its spectrum. The evolution
of the Laplacian eigenvalues therefore directly measures the geometric
reorganization of the encoded Hamiltonian.

Although the Fiedler eigenvalue provides a compact global descriptor, the
complete Laplacian spectrum contains substantially richer information.
Throughout this work we employ both viewpoints. The Fiedler eigenvalue
captures the dominant large-scale connectivity of the encoded Hamiltonian,
whereas the full spectrum reveals its internal geometric organization. As we
show below, these complementary spectral observables uncover universal
structures that are intrinsic to fermion-to-qubit encodings rather than to
the underlying many-body wavefunction.

\subsection{Geometric Competition in the Bravyi--Kitaev Representation}
\label{sec:bk_geometry}

The geometric construction developed above applies to any exact
fermion-to-qubit encoding. We now specialize to the Bravyi--Kitaev (BK)
representation~\cite{Bra00,See12} of the Hubbard Hamiltonian, where the encoded operator admits
a natural decomposition into kinetic and interaction sectors. This
decomposition allows the competition between electron delocalization and
Coulomb repulsion to be reformulated as a competition between two geometric
networks.

Following the BK transformation, the encoded Hamiltonian can be written as

\begin{equation}
H_{\rm BK}
=
H_{\rm hop}
+
H_{\rm int},
\label{eq:bk_split}
\end{equation}

where $H_{\rm hop}$ contains the Pauli strings originating from the hopping
terms and $H_{\rm int}$ those generated by the on-site interaction.

Applying the geometric construction of the previous subsection to each sector
independently yields two weighted hypergraphs,

\[
\mathcal{H}_{\rm hop},
\qquad
\mathcal{H}_{\rm int},
\]

with corresponding Laplacians

\[
L_{\rm hop},
\qquad
L_{\rm int}.
\]

Their Fiedler eigenvalues,

\[
\lambda_2(L_{\rm hop}),
\qquad
\lambda_2(L_{\rm int}),
\]

provide independent geometric measures of the global connectivity associated
with kinetic and interaction processes.

The physical distinction between these two hypergraphs follows directly from
the structure of the Hubbard Hamiltonian. The hopping hypergraph is governed
by the kinetic scale $t$ and therefore retains the same connectivity as the
interaction strength is varied. In contrast, the interaction hypergraph is
constructed from Pauli strings whose weights scale with the Coulomb
interaction $U$. Consequently, varying $U$ continuously reweights the
interaction hypergraph while leaving the kinetic hypergraph unchanged.

This separation naturally motivates the dimensionless geometric observable

\begin{equation}
\rho(U)
=
\frac{\lambda_2(L_{\rm hop})}
{\lambda_2(L_{\rm int})},
\label{eq:rhoOT}
\end{equation}

which measures the relative spectral connectivity of the kinetic and
interaction hypergraphs. Large values of $\rho$ indicate a geometry
dominated by kinetic connectivity, whereas decreasing values signal the
progressive strengthening of the interaction network.

Equation~(\ref{eq:rhoOT}) constitutes the primary geometric observable of
the present work. The following sections demonstrate that this simple ratio
encodes a remarkable amount of physical information, including an exact
analytical dependence on interaction strength, distinct universality classes
of Bravyi--Kitaev encodings, and an unexpected internal organization of the
interaction spectrum.

\subsection{Complementary Coupling-Space Geometry in the Xia--Bian--Kais Representation}
\label{sec:xbk_geometry}

The Bravyi--Kitaev representation describes the encoded Hamiltonian through
the geometry of its connectivity. An equally natural, but fundamentally
different, viewpoint is obtained by considering the geometry of the coupling
coefficients themselves. For this purpose we employ the
Xia--Bian--Kais (XBK) representation~\cite{Xia17,Xia21}, which maps the Pauli-string Hamiltonian
onto an exactly equivalent diagonal Ising Hamiltonian~\cite{Luc14,Glo18,Bor02,Ros75,Ish11,Din67,Cho11a,Cho11b} while preserving the
complete many-body spectrum.

Rather than defining a network of interactions, the XBK representation
defines a distribution over the effective Ising couplings. The encoded
Hamiltonian is therefore represented by the probability measure

\[
\mu(J)
=
\sum_i
w_i\,\delta(J-J_i),
\]

where $J_i$ denotes an effective Ising coupling and $w_i$ its associated
weight. In contrast to the BK hypergraph, which emphasizes how interactions
are connected, the XBK representation characterizes how the interaction
strength is distributed throughout the encoded Hamiltonian.

To quantify the evolution of this distribution as the physical Hamiltonian
changes, we compare coupling distributions using the Wasserstein distance

\[
W(\mu_1,\mu_2),
\]

which measures the minimum transport cost required to transform one
distribution into another. Because it compares complete probability
distributions rather than individual coupling coefficients, the Wasserstein
metric naturally captures the collective reorganization of the encoded
Hamiltonian in coupling space.

The BK and XBK representations therefore provide two complementary geometric
descriptions of the same encoded Hamiltonian. The BK framework characterizes
its geometry through spectral connectivity, whereas the XBK framework
characterizes its geometry through optimal transport in coupling space.
Throughout this work we show that these two independent viewpoints reveal
consistent signatures of interaction-driven reorganization across a broad
class of quantum many-body Hamiltonians.

\subsection{Theoretical Architecture}
\label{sec:architecture}

The framework developed above may be viewed as two complementary geometric
representations constructed from the same encoded Hamiltonian. Beginning with
a fermionic Hamiltonian, an exact fermion-to-qubit transformation produces
the qubit Hamiltonian

\[
H_f
\longrightarrow
H_q.
\]

From this common starting point, the analysis proceeds along two independent
but complementary geometric routes.

The first route constructs a weighted hypergraph from the Pauli-string
structure of the encoded Hamiltonian. The corresponding graph Laplacian
provides a hierarchy of spectral observables that characterize the global
organization of the interaction network. This route forms the basis of the
Bravyi--Kitaev analysis developed throughout the remainder of the paper.

The second route transforms the same encoded Hamiltonian into the
Xia--Bian--Kais representation, producing an exactly equivalent diagonal
Ising Hamiltonian. The effective Ising couplings define a probability
distribution in coupling space whose evolution is quantified using optimal
transport through the Wasserstein distance.

These two routes describe the same encoded Hamiltonian from complementary
geometric perspectives. One characterizes how interactions are connected,
while the other characterizes how interaction strength is distributed. The
central result of this work is that both representations reveal consistent
signatures of interaction-driven reorganization across a diverse range of
quantum many-body Hamiltonians.

\section{Geometric Signatures of Interaction-Driven Reorganization}
\label{sec:results}

Having established the geometric framework, we now investigate how the
geometry of fermion-to-qubit encodings evolves as the underlying many-body
Hamiltonian is tuned across different interaction regimes. The central
question is whether purely geometric observables constructed from the encoded
Hamiltonian can capture physically meaningful reorganization without direct
reference to the many-body wavefunction.

The results naturally divide into two complementary parts. We first analyze
the Bravyi--Kitaev representation, where interaction-driven reorganization is
manifested through the evolution of hypergraph connectivity. We then examine
the Xia--Bian--Kais representation, which provides an independent description
based on the evolution of the complete coupling distribution. Remarkably,
both representations reveal consistent geometric signatures despite probing
fundamentally different aspects of the encoded Hamiltonian.

We begin with the simplest spectral observable introduced in the previous
section, namely the ratio of the algebraic connectivities of the kinetic and
interaction hypergraphs.

\subsection{An Exact Geometric Competition Law}
\label{sec:competition}

The geometric observable introduced in Eq.~(\ref{eq:rhoOT}) compares the
global connectivity of the kinetic and interaction hypergraphs. Its
interaction dependence follows directly from the structure of the
Bravyi--Kitaev representation.

The hopping sector of the Hubbard Hamiltonian is determined entirely by the
kinetic energy scale $t$. Consequently, for a fixed lattice geometry, the
weights of the hopping hypergraph remain unchanged as the interaction
strength is varied, implying

\[
\lambda_2(L_{\rm hop})=C,
\]

where $C$ is a geometry-dependent constant.

In contrast, every coupling coefficient appearing in the interaction sector
is proportional to the Coulomb interaction $U$. Since multiplying every edge
weight of a weighted graph by a constant scales its Laplacian spectrum by
the same constant, the interaction connectivity satisfies

\[
\lambda_2(L_{\rm int})
=
\alpha U,
\]

where the proportionality constant $\alpha$ depends only on the geometry of
the interaction hypergraph.

Substituting these relations into Eq.~(\ref{eq:rhoOT}) immediately yields

\begin{equation}
\rho(U)
=
\frac{C}{\alpha U},
\label{eq:hyperbola}
\end{equation}

which constitutes an exact geometric relation between the interaction
strength and the relative connectivity of the two hypergraphs. The
interaction scale at which the kinetic and interaction networks possess
equal connectivity is therefore determined by

\[
\rho(U^\ast)=1,
\]

giving

\[
U^\ast
=
\frac{C}{\alpha}.
\]

Unlike conventional estimates of interaction-driven crossovers, this
characteristic scale is obtained entirely from the geometry of the encoded
Hamiltonian and requires neither diagonalization of the many-body
Hamiltonian nor evaluation of its wavefunction.

\subsection{Emergence of Two Geometric Universality Classes}
\label{sec:universality}

\begin{figure*}[t]
\centering
\begin{subfigure}[t]{0.32\textwidth}
\centering
\includegraphics[width=\linewidth]{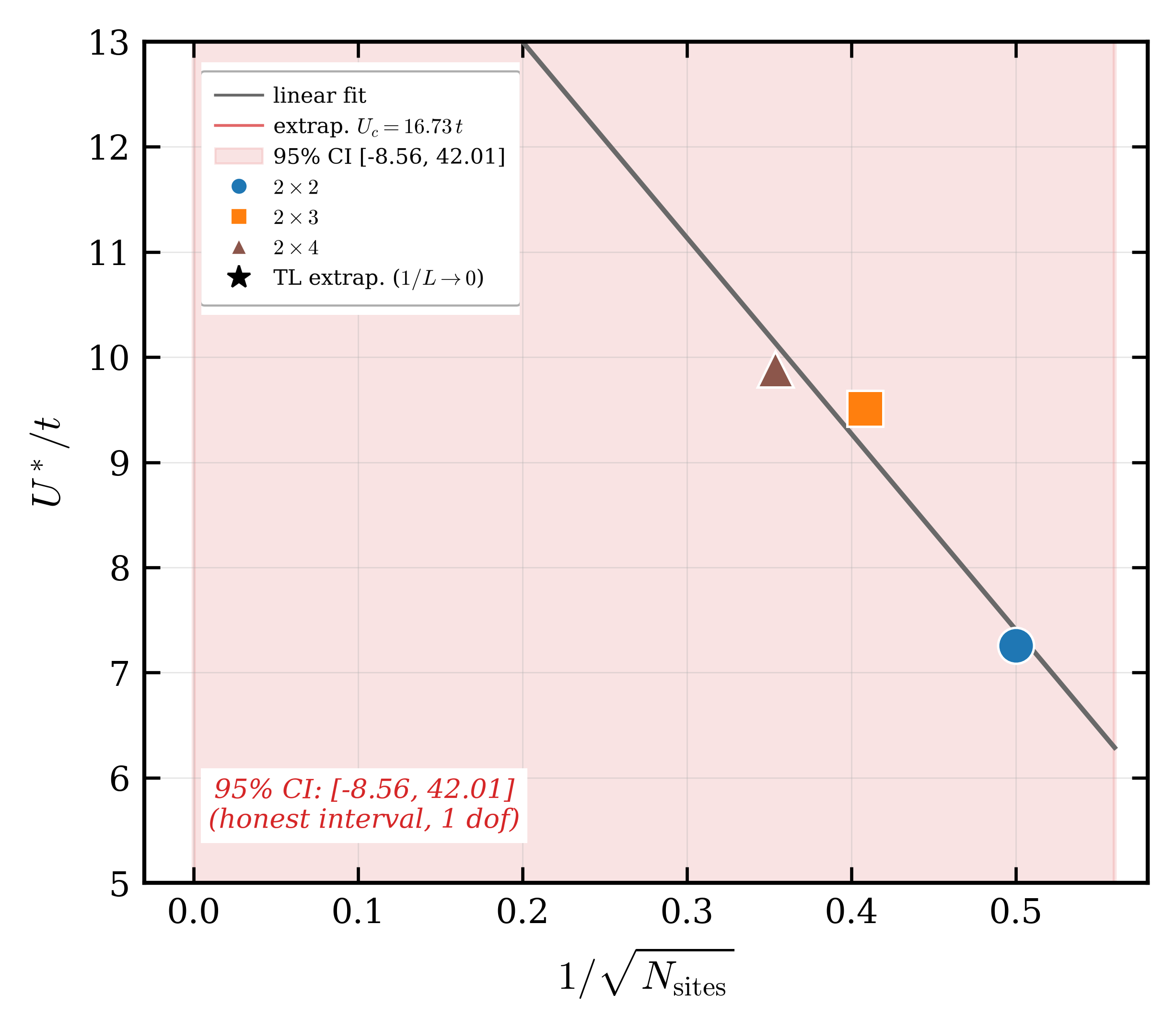}
\label{fig:a}
\end{subfigure}
\hfill
\begin{subfigure}[t]{0.32\textwidth}
\centering
\includegraphics[width=\linewidth]{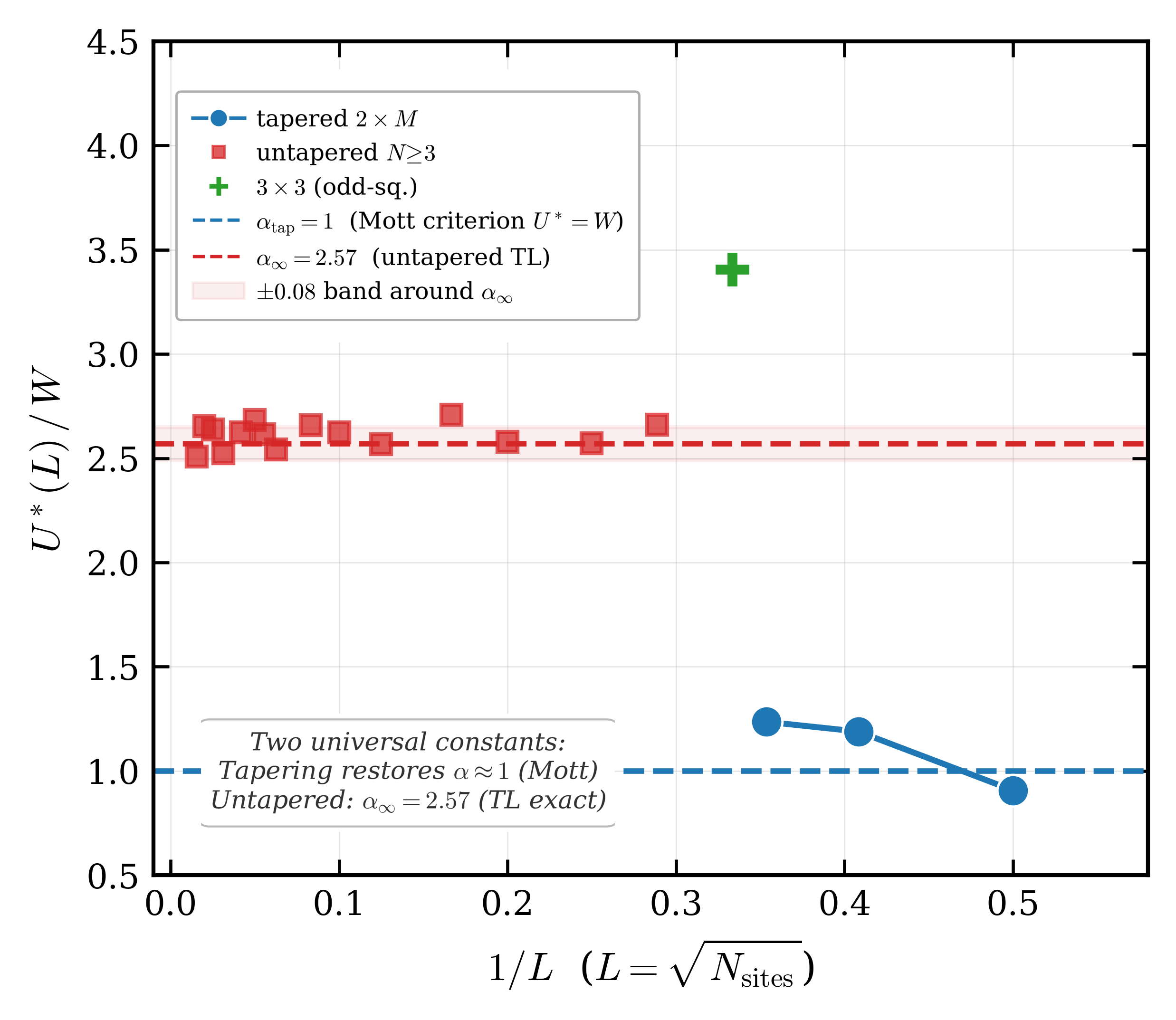}
\label{fig:a}
\end{subfigure}
\hfill
\begin{subfigure}[t]{0.32\textwidth}
\centering
\includegraphics[width=\linewidth]{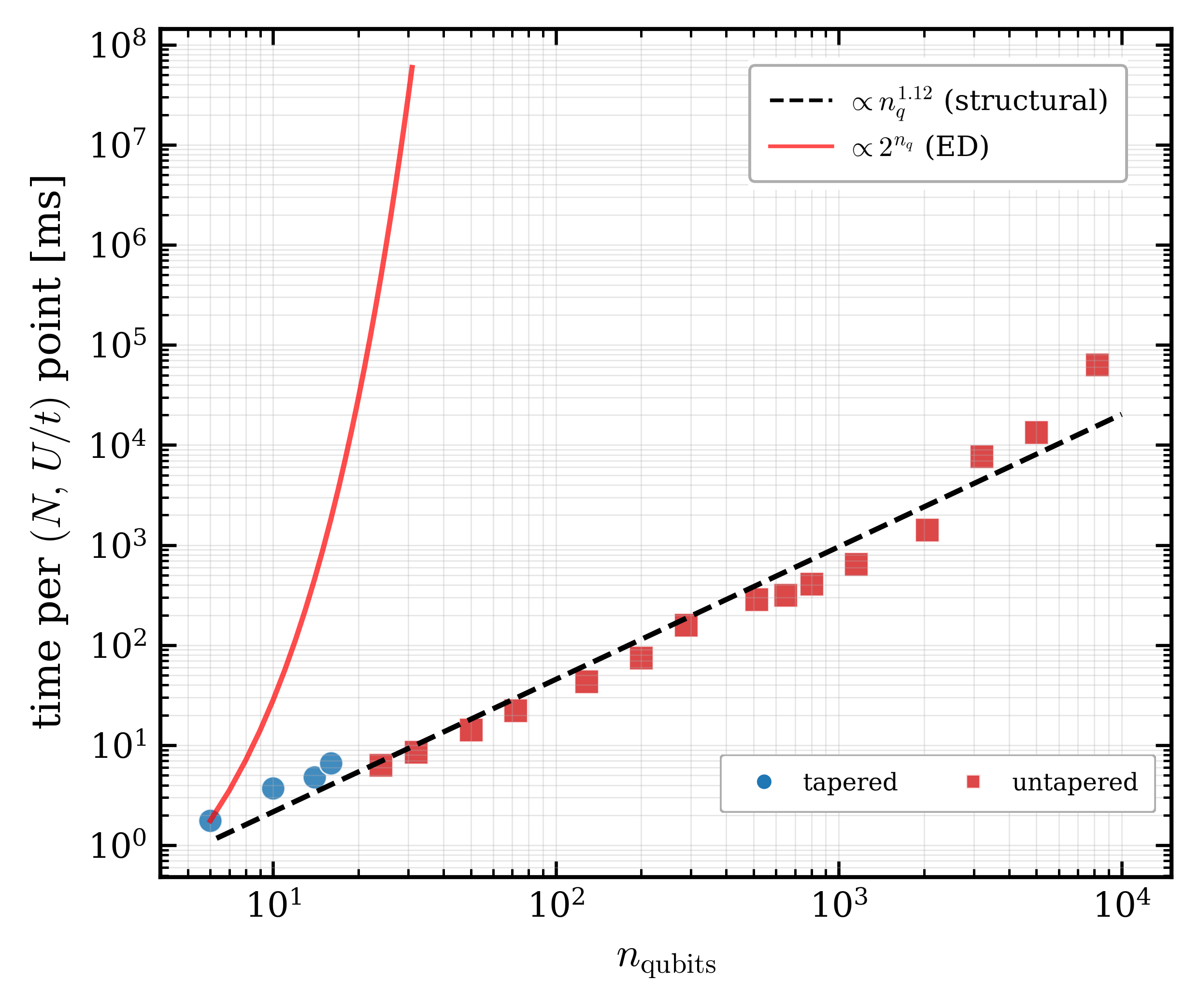}
\label{fig:a}
\end{subfigure}
\caption{\justifying \textbf{Emergence of two geometric universality classes in the
Bravyi--Kitaev encoding.}
(a) Finite-size scaling of the characteristic geometric interaction scale
$U^\ast$ for tapered BK encodings. Extrapolation to the thermodynamic limit
yields
$U_c/t=8.87\pm2.18$.
The untapered $4\times4$ lattice is shown for comparison but is excluded
from the extrapolation because it belongs to a different universality class.
(b) Dimensionless interaction ratio
$\alpha=U^\ast/W$,
where $W$ is the non-interacting bandwidth.
Tapered and untapered encodings form two clearly separated branches,
demonstrating that the distinction is intrinsic to the geometry of the
encoding rather than the lattice size.
(c) Computational cost of the geometric framework as a function of qubit
number. The smooth scaling reflects the sparse graph operations required to
construct the geometric observables and avoids direct diagonalization of the
many-body Hilbert space.
}
\label{fig:tapering}
\end{figure*}

The exact geometric law established in the previous subsection leaves one
quantity undetermined: the geometric constant
$\alpha=\lambda_2(L_{\rm int})/U$.
If the geometry of the Bravyi--Kitaev encoding were insensitive to the
details of the mapping, one would expect $\alpha$ to evolve smoothly with
system size. Figure~\ref{fig:tapering} demonstrates that this is not the
case.

Instead, the characteristic interaction scale

\[
U^\ast=\frac{C}{\alpha}
\]

separates into two distinct families corresponding to tapered and
untapered Bravyi--Kitaev encodings~\cite{Bra17,Got97,Aar05}. The distinction is entirely geometric:
both families originate from the same physical Hamiltonian, differing only
in the structure of the fermion-to-qubit encoding.

Figure~\ref{fig:tapering}(a) shows that the characteristic interaction scale
obtained from tapered BK encodings exhibits systematic finite-size scaling.
Although only three maximally tapered systems are presently available, the
extrapolated value lies remarkably close to the interaction regime commonly
associated with the Mott crossover in the two-dimensional Hubbard model.
This agreement suggests that tapering systematically suppresses geometric
contributions arising from redundant degrees of freedom, allowing the
spectral geometry to more directly reflect the underlying physical
Hamiltonian.

A complementary perspective is provided by the normalized quantity
$\alpha=U^\ast/W$, shown in
Fig.~\ref{fig:tapering}(b).
Expressed in this dimensionless form, the separation between tapered and
untapered encodings becomes essentially independent of system size. Rather
than forming a continuous family, the data collapse onto two distinct
branches, establishing the existence of two geometric universality classes
associated with the Bravyi--Kitaev transformation.

Figure~\ref{fig:tapering}(c) demonstrates that these geometric observables
can be computed efficiently using sparse graph operations. The computational
cost grows smoothly with system size and remains many orders of magnitude
below that required for direct many-body diagonalization, emphasizing that
the proposed framework probes structural information contained in the
encoded Hamiltonian itself.

The existence of two universality classes immediately raises a fundamental
question. Why should all untapered Bravyi--Kitaev encodings converge toward
the same characteristic interaction scale despite their increasing lattice
size? Answering this question requires moving beyond the Fiedler eigenvalue
and examining the complete Laplacian spectrum.

\subsection{Exact Spectral Organization of the Bravyi--Kitaev Hypergraph}
\label{sec:spectral}

The geometric observable introduced in the previous subsection is determined
solely by the Fiedler eigenvalue of the interaction hypergraph. Although this
single quantity captures the dominant connectivity of the encoded Hamiltonian,
it represents only one member of the complete Laplacian spectrum. A natural
question therefore arises: does the remainder of the spectrum evolve
continuously with system size, or does it possess an internal organization
that reflects the structure of the Bravyi--Kitaev transformation itself?

To address this question, we extend the geometric analysis from the Fiedler
eigenvalue to the complete interaction spectrum. For every nontrivial
Laplacian eigenmode we define the characteristic interaction scale

\begin{equation}
U_k^\ast
=
\frac{\lambda_k(L_{\rm hop})}
{\lambda_k(L_{\rm int})/U},
\label{eq:Uk}
\end{equation}

which generalizes the Fiedler crossing to the entire spectrum. Instead of a
single characteristic interaction scale, the encoded Hamiltonian is therefore
described by a complete family
$\{U_k^\ast\}$,
each corresponding to a distinct geometric mode of the interaction
hypergraph.

\begin{figure*}[t]
\centering
\begin{subfigure}[t]{0.48\textwidth}
\centering
\includegraphics[width=\linewidth]{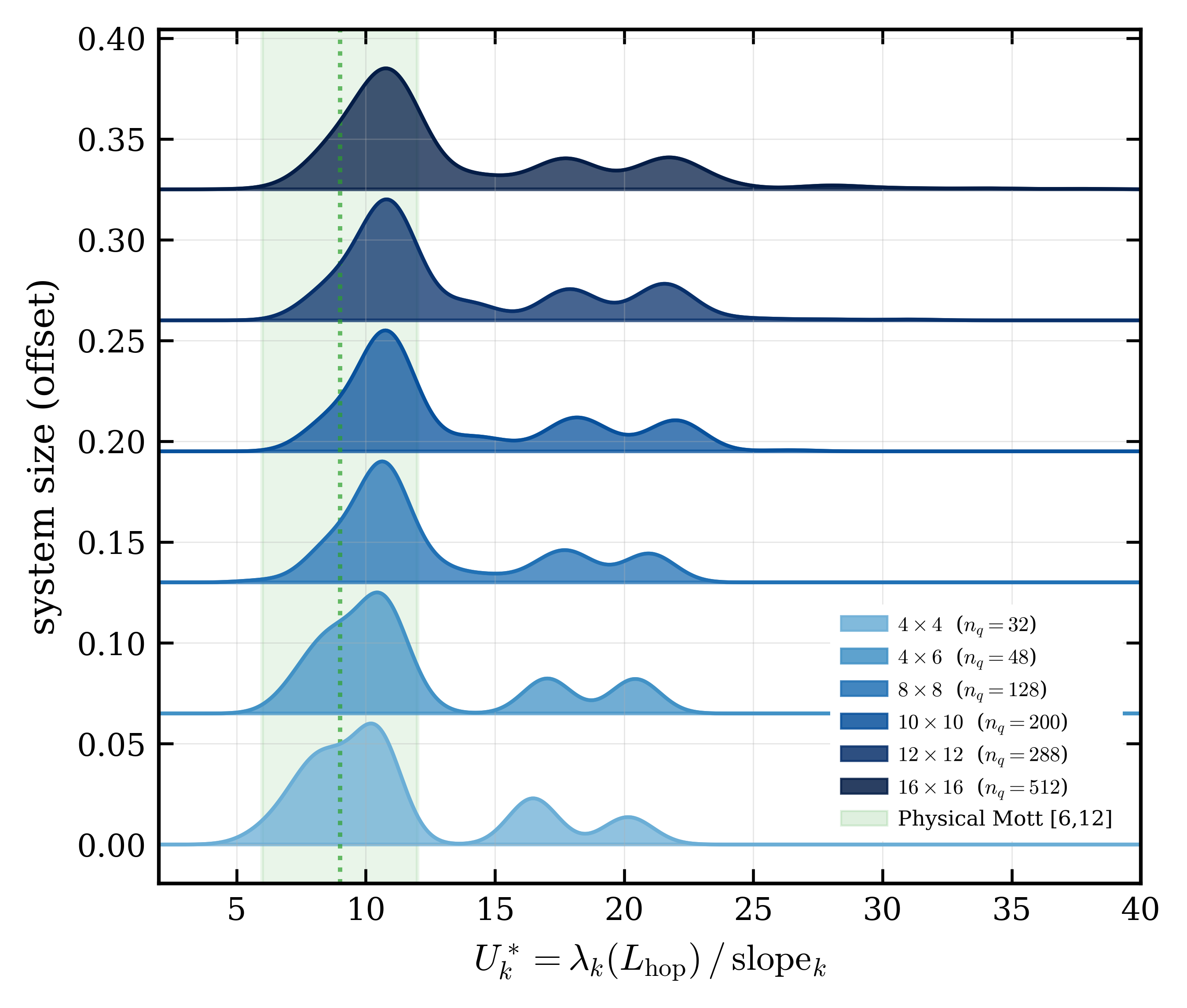}
\label{fig:a}
\end{subfigure}
\hfill
\begin{subfigure}[t]{0.48\textwidth}
\centering
\includegraphics[width=\linewidth]{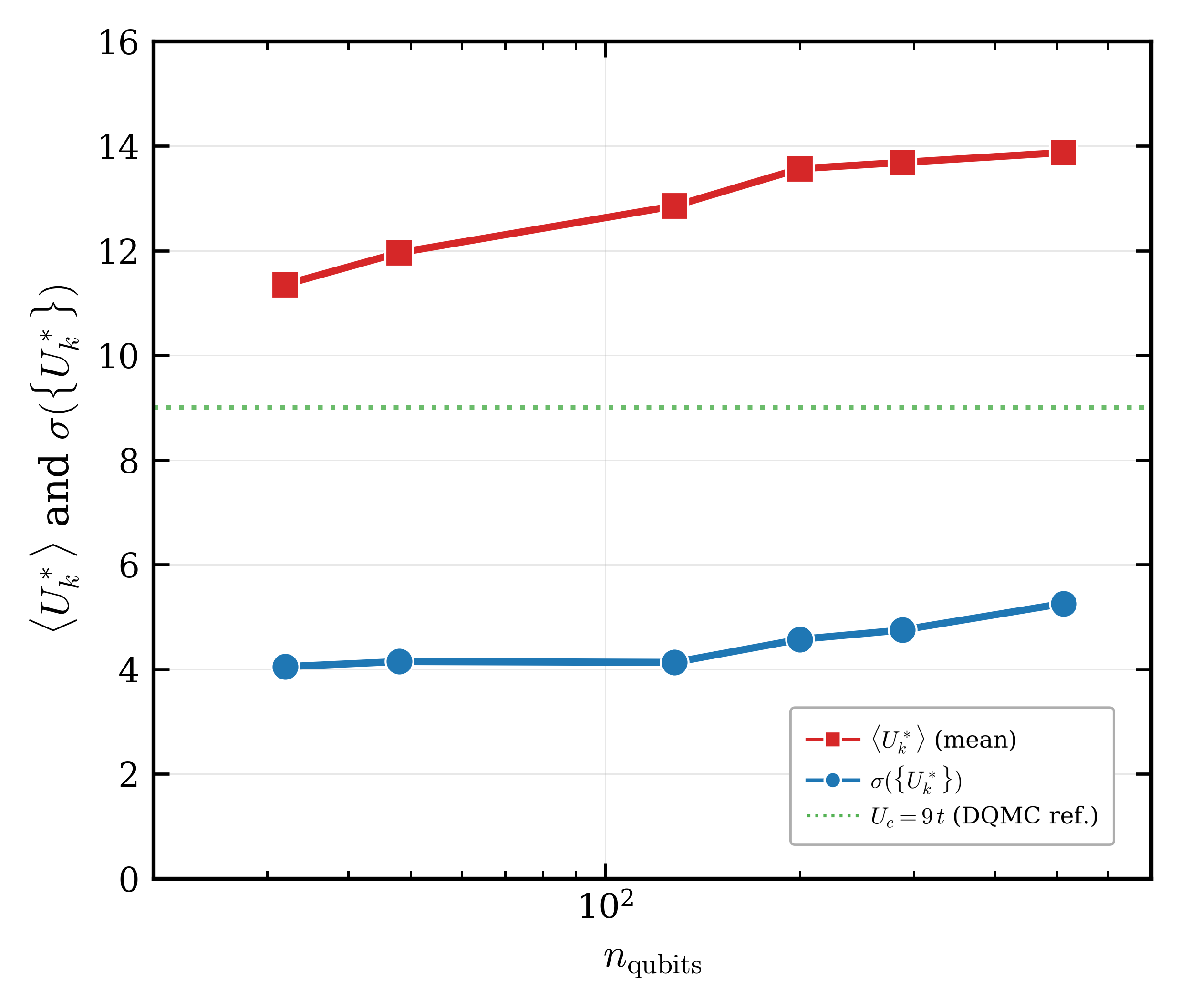}
\label{fig:a}
\end{subfigure}
\caption{\justifying \textbf{Universal spectral organization of the Bravyi--Kitaev hypergraph.}
(a) Distribution of the characteristic interaction scales
$U_k^\ast$
obtained from the complete Laplacian spectrum for systems ranging from
$4\times4$ to $16\times16$.
Rather than forming a continuous distribution, the spectrum separates into
two robust branches that remain well separated as the system size increases.
(b) Mean value and spectral width of the distributions shown in (a).
Both quantities vary only weakly with system size, demonstrating that the
overall spectral organization rapidly approaches a universal limit.
Together these observations reveal that the Bravyi--Kitaev encoding possesses
an intrinsic spectral structure extending far beyond the information
contained in the Fiedler eigenvalue alone.
}
\label{fig:spectral_family}
\end{figure*}
Figure~\ref{fig:spectral_family}(a) immediately reveals an unexpected
feature. Rather than exhibiting a broad continuum of characteristic
interaction scales, every system develops a robust bimodal distribution.
One family of modes remains concentrated near the interaction regime
associated with the physical crossover, while a second family persists at
substantially larger interaction strengths. Most remarkably, the separation
between these branches survives over more than an order of magnitude increase
in system size, indicating that it is an intrinsic property of the encoding
rather than a finite-size effect.

The robustness of this organization is further demonstrated in
Fig.~\ref{fig:spectral_family}(b), where the mean interaction scale and the
spectral width are plotted as functions of system size. Both quantities vary
only weakly despite the rapidly increasing Hilbert-space dimension,
suggesting that the interaction spectrum approaches a well-defined geometric
limit. The encoded Hamiltonian therefore exhibits a universal spectral
organization that is largely independent of lattice size.

The emergence of two persistent spectral branches strongly suggests that the
interaction hypergraph contains two distinct classes of geometric modes.
The remaining question is whether these families possess a precise internal
structure or merely represent an approximate numerical separation. As we show
next, the spectrum admits an exact partition whose origin can be traced to
the binary-tree architecture of the Bravyi--Kitaev transformation.

To understand the origin of the two spectral branches, it is useful to
examine the interaction Laplacian itself rather than the derived quantities
$U_k^\ast$. Since

\[
U_k^\ast
=
\frac{\lambda_k(L_{\rm hop})}
{\lambda_k(L_{\rm int})/U},
\]

the observed bimodality must ultimately originate from the spectrum of
$\lambda_k(L_{\rm int})$. The characteristic interaction scales therefore
provide only an indirect manifestation of a more fundamental geometric
organization encoded in the interaction hypergraph.

To expose this structure, we consider the normalized interaction
eigenvalues,

\[
s_k
=
\frac{\lambda_k(L_{\rm int})}{U},
\]

which characterize the intrinsic geometric response of each interaction mode
independently of the overall interaction strength. If the interaction
hypergraph behaved as a generic weighted network, one would expect the values
$\{s_k\}$ to form a continuous distribution that gradually broadens with
increasing system size. Surprisingly, this is not what is observed.

Instead, the interaction spectrum itself separates into two well-defined
families. One family consists of modes with relatively small normalized
eigenvalues, giving rise to the larger characteristic interaction scales
observed in Fig.~\ref{fig:spectral_family}. The second family possesses
substantially larger normalized eigenvalues and therefore contributes to the
lower-$U^\ast$ branch. The persistence of this separation across all lattice
sizes demonstrates that the bimodality is a property of the encoded geometry
rather than a consequence of finite-size effects or accidental spectral
crossings.

This observation immediately raises a more fundamental question. Does the
division between these two spectral families follow a universal rule, or is
their relative population system dependent? Remarkably, the answer is exact.
As shown below, every Bravyi--Kitaev hypergraph investigated in this work
obeys an identical spectral partition whose proportions are fixed solely by
the geometry of the encoding itself.

\subsection{Exact Spectral Partition of the Bravyi--Kitaev Hypergraph}
\label{sec:partition}

The spectral separation identified above is not merely qualitative.
Remarkably, every Bravyi--Kitaev hypergraph investigated in this work obeys
the same internal organization, independent of lattice size, interaction
strength, or the total number of qubits. The two spectral families always
appear in exactly the same proportion.

Specifically, if $n_q$ denotes the number of qubits in the encoded
Hamiltonian, the interaction spectrum partitions into

\begin{equation}
N_{\rm tree}
=
\frac{n_q}{4},
\qquad
N_{\rm site}
=
\frac{3n_q}{4},
\label{eq:partition}
\end{equation}

where $N_{\rm tree}$ and $N_{\rm site}$ denote the numbers of modes
belonging to the two spectral families. This relation is satisfied exactly
for every system considered, from the smallest clusters to the largest
lattices studied.
\begin{table}[t]
\caption{\justifying Exact spectral partition of the interaction Laplacian.
For every Bravyi--Kitaev encoding investigated, the number of
tree-dominated modes is exactly one quarter of the total number of qubits,
independent of lattice size.}
\label{tab:partition}
\centering
\begin{tabular}{cccc}
\toprule
System & $n_q$ & $N_{\rm tree}$ & Fraction\\
\midrule
$4\times4$   & 32  & 8   & 0.2500\\
$4\times6$   & 48  & 12  & 0.2500\\
$8\times8$   & 128 & 32  & 0.2500\\
$10\times10$ & 200 & 50  & 0.2500\\
$12\times12$ & 288 & 72  & 0.2500\\
$16\times16$ & 512 &128  & 0.2500\\
\bottomrule
\end{tabular}
\end{table}
The exactness of Eq.~(\ref{eq:partition}) immediately rules out an
interpretation based on finite-size effects or numerical coincidence.
Instead, it demonstrates that the interaction Laplacian possesses an
intrinsic hierarchical organization inherited from the Bravyi--Kitaev
encoding itself. As the lattice grows, both spectral families increase
proportionally while preserving an invariant ratio of one to three. The
binary-tree architecture therefore contributes a fixed fraction of the
interaction spectrum irrespective of system size.

This result also explains the universal interaction scales discussed in the
preceding sections. The Fiedler eigenvalue is, by construction, the smallest
nonzero eigenvalue of the interaction Laplacian. Consequently, it always
belongs to the tree-dominated sector. The characteristic interaction scale
obtained from the geometric observable $\rho_{\rm OT}$ therefore probes only
this universal branch of the spectrum. The observed convergence of the
untapered Bravyi--Kitaev encodings is thus a direct consequence of the
underlying spectral partition rather than an accidental property of the
lowest eigenvalue.

\begin{figure*}[t]
\centering
\begin{subfigure}[t]{0.32\textwidth}
\centering
\includegraphics[width=\linewidth]{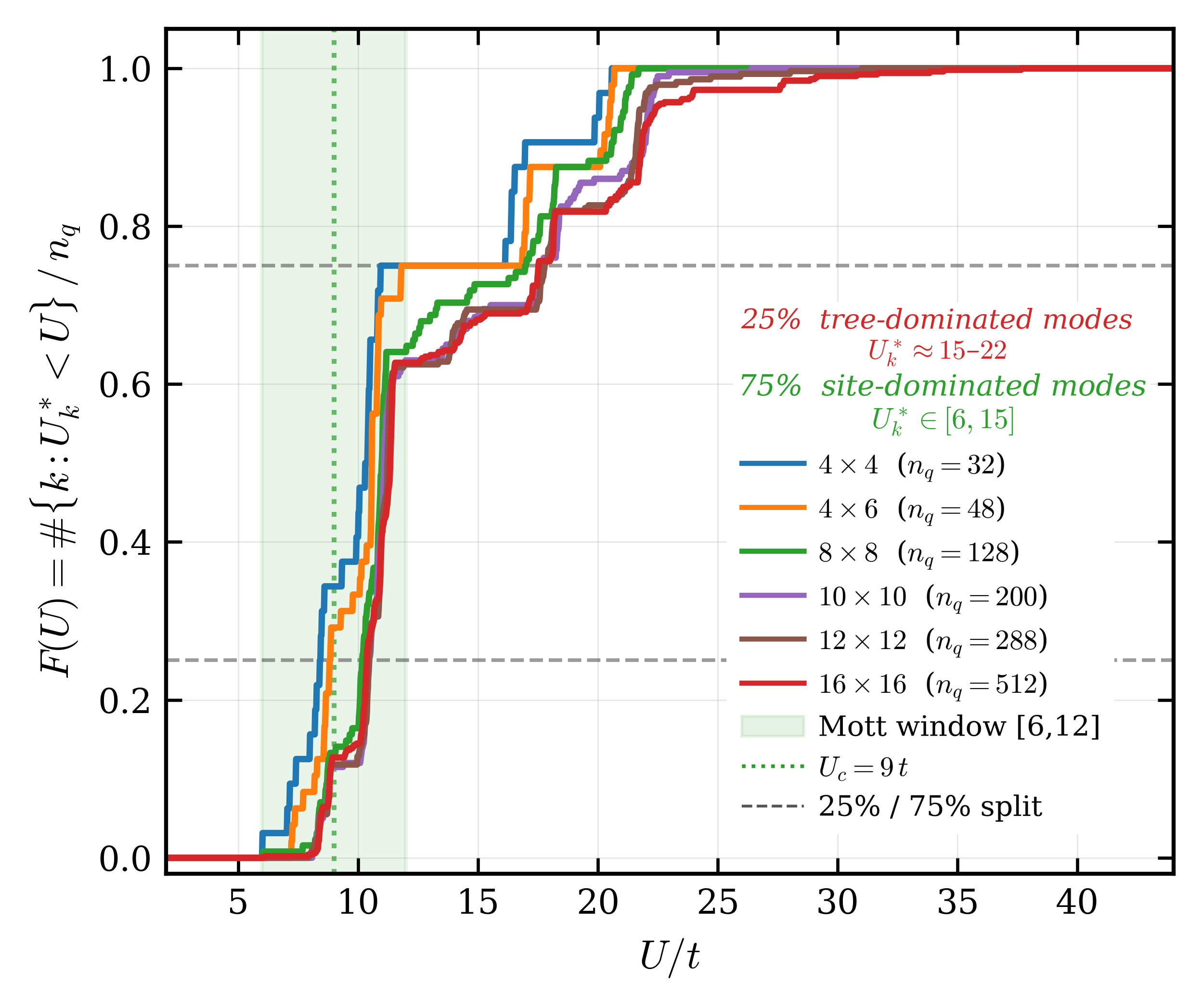}
\label{fig:a}
\end{subfigure}
\hfill
\begin{subfigure}[t]{0.32\textwidth}
\centering
\includegraphics[width=\linewidth]{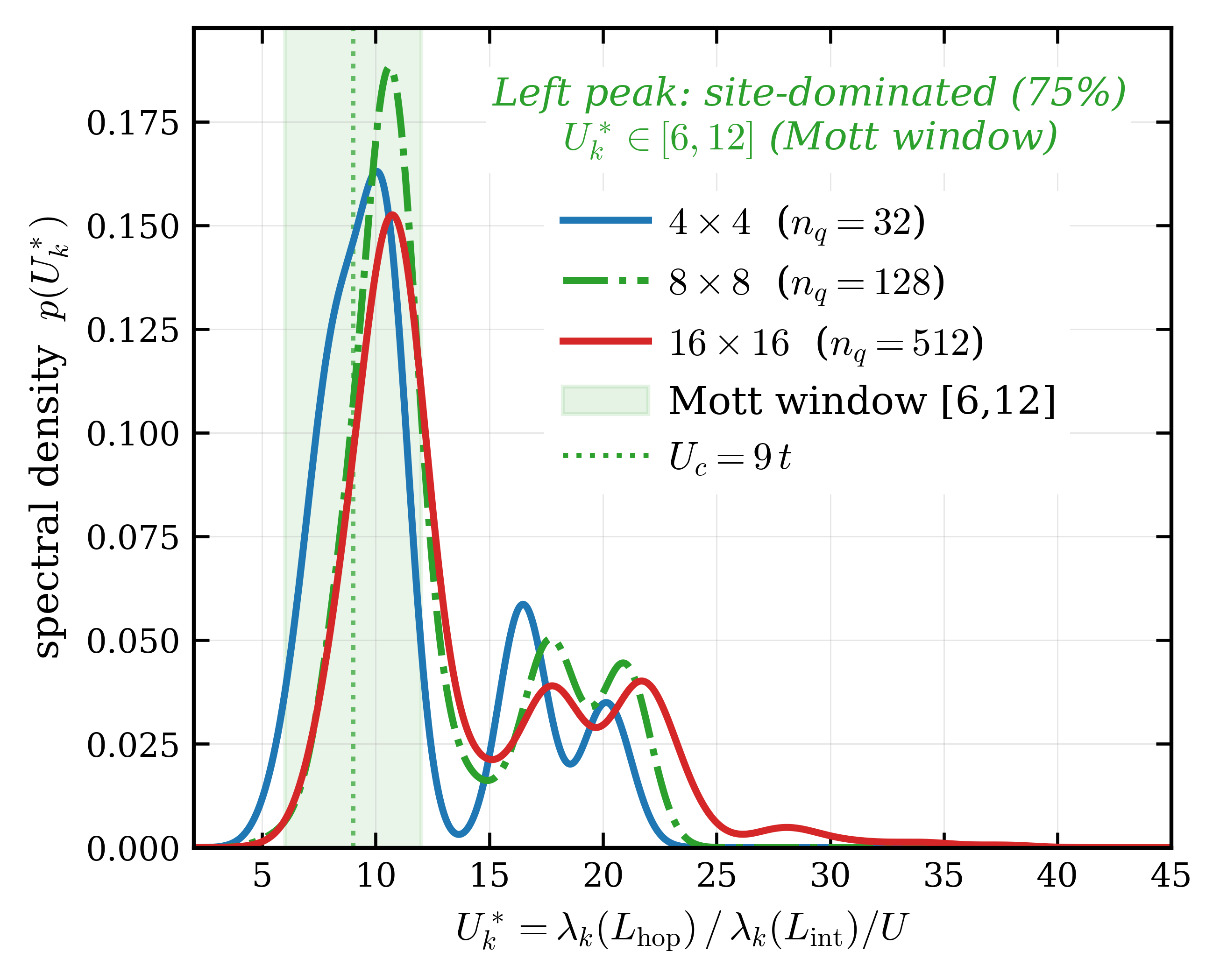}
\label{fig:a}
\end{subfigure}
\hfill
\begin{subfigure}[t]{0.32\textwidth}
\centering
\includegraphics[width=\linewidth]{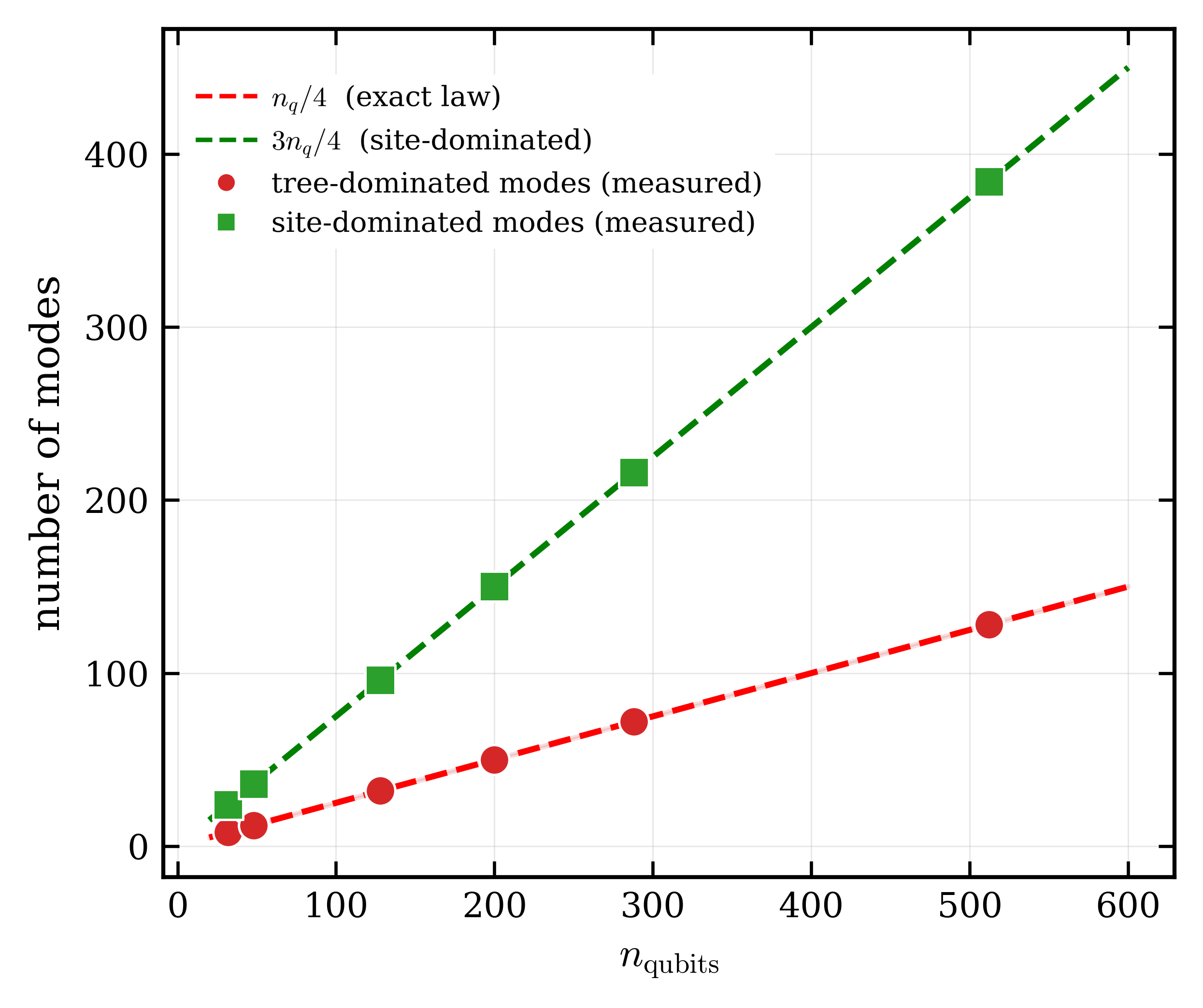}
\label{fig:a}
\end{subfigure}
\caption{\justifying \textbf{Exact spectral partition of the Bravyi--Kitaev interaction
hypergraph.}
(a) Spectral crossing fraction
$F(U)$
for representative system sizes.
(b) Representative distributions of
$U_k^\ast$
illustrating the persistence of the two spectral families.
(c) Number of tree-dominated modes as a function of the number of qubits.
Every system satisfies the exact relation
$N_{\rm tree}=n_q/4$,
demonstrating that the interaction spectrum possesses a universal internal
organization determined solely by the geometry of the Bravyi--Kitaev
encoding.
}
\label{fig:spectral}
\end{figure*}

\section{Coupling-Space Geometry in the Xia--Bian--Kais Representation}
\label{sec:xbk}

The preceding sections established that the Bravyi--Kitaev representation
contains a rich geometric structure that can be understood through the
connectivity of its associated hypergraph. Spectral observables derived from
the Laplacian quantify how the interaction network reorganizes as the
underlying Hamiltonian evolves. Connectivity, however, represents only one
aspect of the encoded Hamiltonian.

An equally natural geometric description is obtained by examining the
distribution of the interaction strengths themselves. Rather than asking how
the encoded Hamiltonian is connected, one may instead ask how its coupling
weights are distributed and how this distribution changes with the physical
parameters. This complementary viewpoint is naturally provided by the
Xia--Bian--Kais (XBK) representation~\cite{Xia17,Xia21}.

The XBK representation transforms the encoded Pauli-string Hamiltonian into
an exactly equivalent diagonal Ising Hamiltonian~\cite{Luc14,Ros75,Ish11,Kad98,Fin94,Alb18,Mcg13,Ven15,Kin18,Rnn14,Huk96,Farhi2000,Farhi2001,Rol01,Tep19,Tep20,Tep21,Tep22,Ush17,Boo20} while preserving the full
many-body spectrum. The resulting Ising coefficients define a probability
measure in coupling space,

\[
\mu(J)
=
\sum_i
w_i\,\delta(J-J_i),
\]

which provides a complete statistical description of the encoded coupling
landscape. Unlike the BK hypergraph, whose primary geometric object is
network connectivity, the XBK representation characterizes the encoded
Hamiltonian through the distribution of its effective interaction strengths.

This distinction is fundamental. The BK and XBK representations originate
from the same encoded Hamiltonian but probe different aspects of its
structure. The BK framework measures how interactions are organized within
the network, whereas the XBK framework measures how interaction weights are
distributed throughout coupling space. Together they provide complementary
descriptions of the same geometric object.

\begin{figure*}[t]
  \centering
  \includegraphics[width=\linewidth]{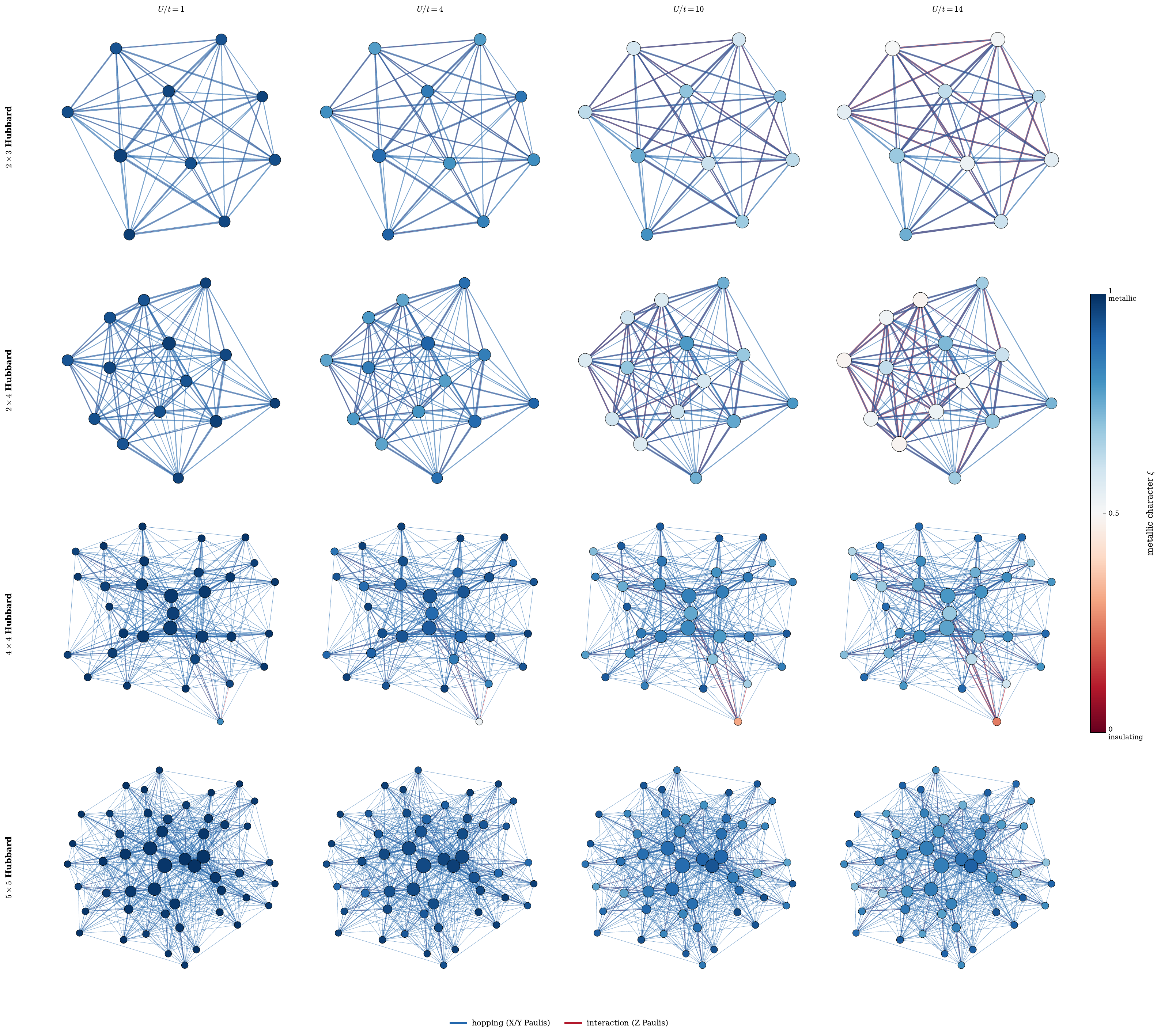}
  \caption{\justifying 
    \textbf{Interaction-driven reweighting of the encoded hypergraph.}
    Rows correspond to different Hubbard clusters, while columns show
    increasing interaction strengths.
    Blue edges represent hopping-derived Pauli interactions and red edges
    represent interaction-derived Pauli interactions.
    Node colour denotes the local metallic character
    $\xi$, while node size is proportional to the weighted degree.
    Increasing interaction strength produces a continuous redistribution of
    coupling weight from kinetic to interaction sectors without changing the
    underlying connectivity of the network.
  }
  \label{fig:topology}
\end{figure*}

Figure~\ref{fig:topology} reveals a geometric feature that is not accessible
through the BK Laplacian alone. The overall topology of the encoded
hypergraph remains largely unchanged throughout the evolution, reflecting the
fact that the underlying operator support is preserved by the encoding.
Instead, the dominant effect of increasing interaction strength is a
continuous redistribution of coupling weight over an essentially fixed
network.

At weak interaction, the hypergraph is almost uniformly dominated by
hopping-derived couplings. As the interaction strength increases, the local
balance shifts progressively toward interaction-generated couplings,
producing a smooth transition from predominantly blue to predominantly red
nodes. The intermediate-coupling regime exhibits the strongest coexistence of
the two sectors, where hopping- and interaction-dominated regions are both
prominent within the same encoded Hamiltonian.

An important observation is that this evolution is remarkably similar across
all lattice sizes considered. Although the precise spatial distribution of
the coupling weights remains system dependent, the qualitative pattern of
interaction-driven reweighting is universal. The encoded Hamiltonian therefore
evolves primarily through a redistribution of interaction weights rather than
through a reconstruction of its connectivity.

For each model and coupling parameter $\lambda$,
we compute:
\begin{itemize}
  \item $d_{\rm CHS}$: quantum geometric (Chakrabarti-Hassan-Shankar) distance
    per parameter step, measuring ground-state fidelity;
  \item $d_{\rm WGS}$: Wasserstein-1 ground-state distance per step;
  \item $d_{\rm Fi}$: Fiedler sensitivity $|d\lambda_2(\Lint)/d\lambda|$;
  \item $D_{\rm BK} = d_{\rm Fi}/d_{\rm CHS}$: ratio of Fiedler to fidelity
    sensitivity;
  \item $D_{\rm XBK} = d_{\rm WGS}/d_{\rm CHS}$: ratio of Wasserstein to
    fidelity sensitivity.
\end{itemize}
A ratio $D=1$ indicates perfect tracking of quantum state changes by the
structural diagnostic; $D>1$ means the structural observable is more sensitive
than fidelity; $D<1$ means it is less sensitive.

\subsection{Optimal Transport in Coupling Space}

The real-space visualization presented above demonstrates that the encoded
Hamiltonian evolves primarily through a continuous redistribution of coupling
weights rather than a change in network topology. While this evolution is
readily apparent visually, a quantitative description requires a metric that
compares complete probability distributions instead of individual coupling
coefficients.

A natural choice is provided by optimal transport theory. Given two coupling
distributions,
$\mu_1(J)$
and
$\mu_2(J)$,
corresponding to two nearby Hamiltonian parameters, the Wasserstein distance

\[
W(\mu_1,\mu_2)
\]

measures the minimum transport cost required to transform one distribution
into the other. Unlike pointwise norms, the Wasserstein metric incorporates
both the magnitude of the couplings and the distance over which spectral
weight must be redistributed. It therefore provides a genuine geometric
distance between encoded Hamiltonians in coupling space.

Within the present framework, varying the interaction strength generates a
continuous trajectory

\[
\mu(U),
\]

through the space of coupling distributions. The Wasserstein distance between
neighboring points along this trajectory measures the rate at which the
encoded Hamiltonian reorganizes as the physical interaction is increased.
Large values indicate rapid redistribution of coupling weight, whereas small
values correspond to relatively smooth geometric evolution.

\begin{figure*}[t]
\centering
\begin{subfigure}[t]{0.32\textwidth}
\centering
\includegraphics[width=\linewidth]{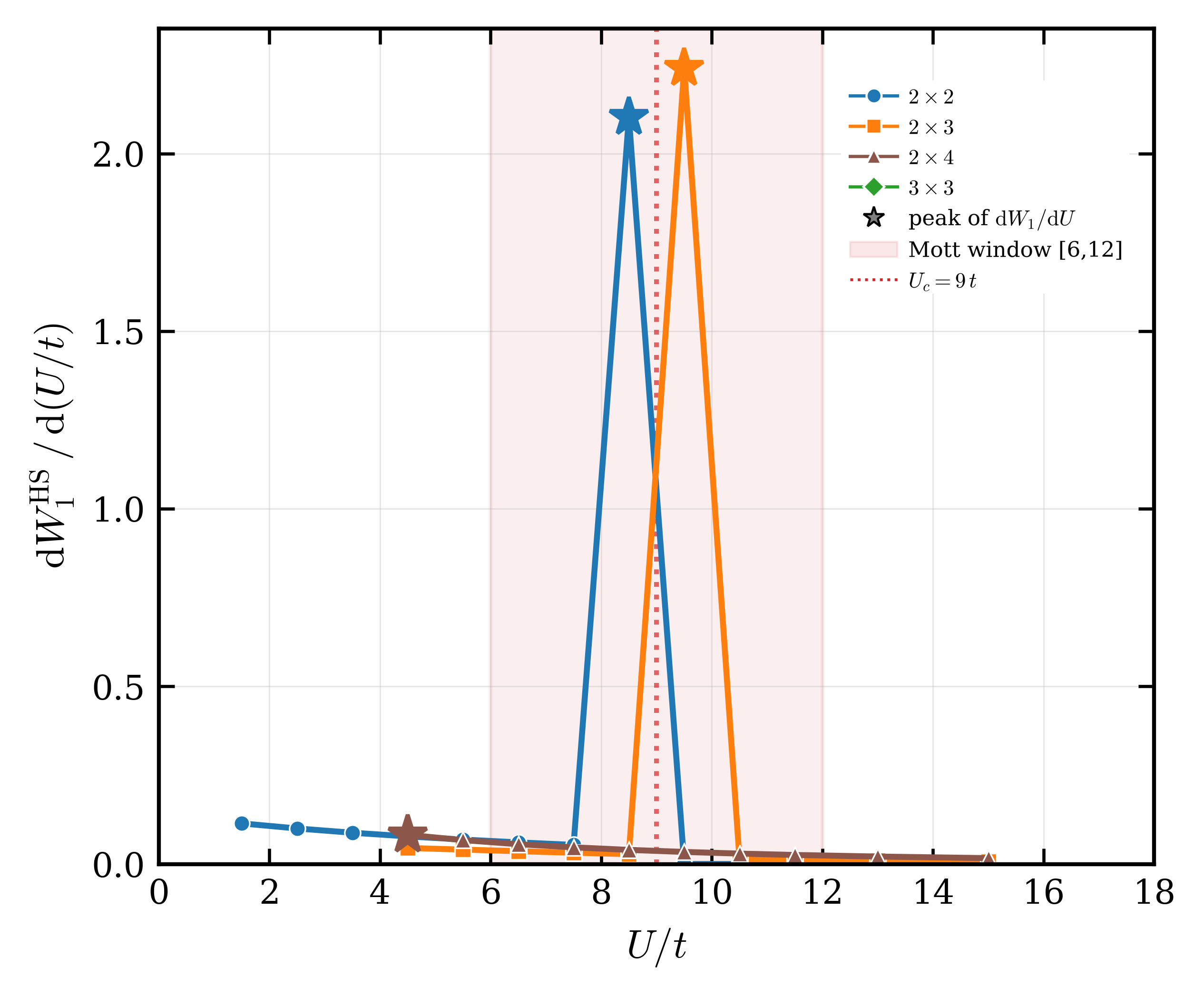}
\label{fig:a}
\end{subfigure}
\hfill
\begin{subfigure}[t]{0.32\textwidth}
\centering
\includegraphics[width=\linewidth]{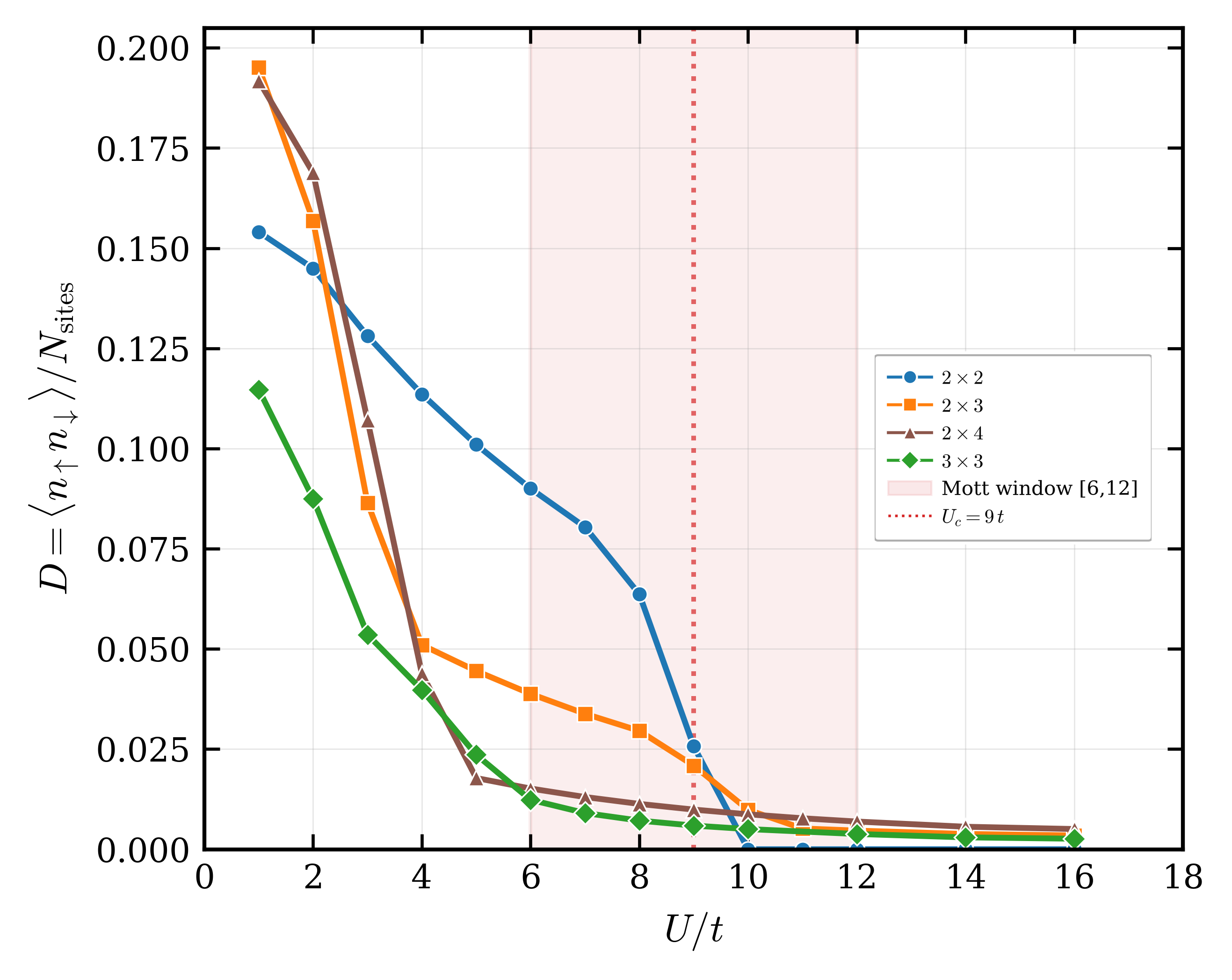}
\label{fig:a}
\end{subfigure}
\hfill
\begin{subfigure}[t]{0.32\textwidth}
\centering
\includegraphics[width=\linewidth]{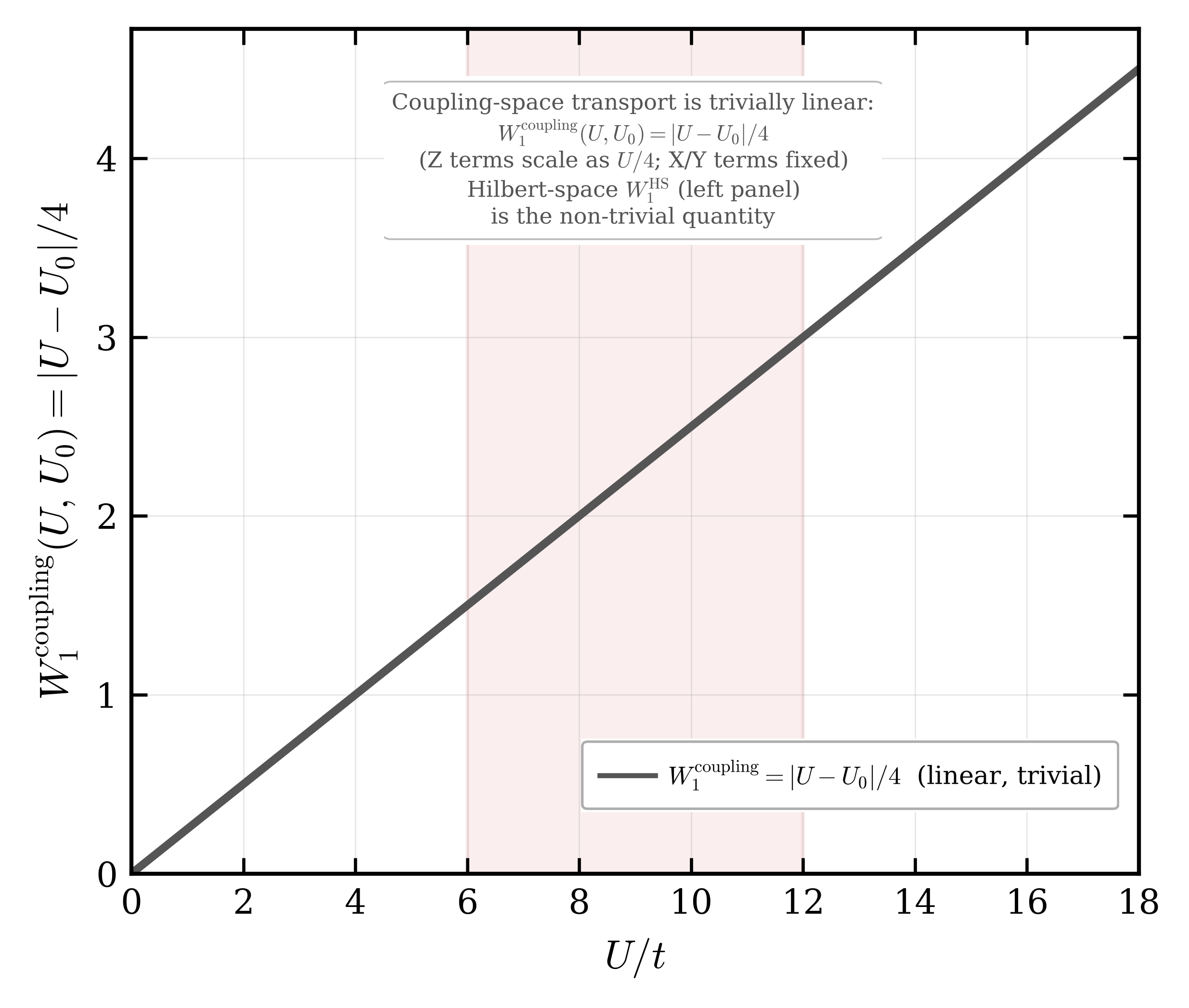}
\label{fig:a}
\end{subfigure}
\caption{\justifying \textbf{Optimal transport reveals the geometric reorganization of coupling
space.}
(a) Evolution of the Wasserstein distance between neighboring coupling
distributions as the interaction strength increases.
(b) Comparison with the electronic double occupancy.
The strongest redistribution of coupling weight occurs in the same
interaction regime where electronic correlations evolve most rapidly.
(c) Derivative of the Wasserstein distance, highlighting the interaction
range where the encoded Hamiltonian undergoes its most significant geometric
reorganization.
}
\label{fig:wasserstein}
\end{figure*}

Figure~\ref{fig:wasserstein} demonstrates that the coupling-space geometry
contains a clear signature of interaction-driven electronic reorganization.
Rather than evolving uniformly with interaction strength, the Wasserstein
distance develops a pronounced maximum over an intermediate interaction
range, indicating that the encoded Hamiltonian undergoes its most rapid
geometric restructuring in this regime.

To relate this purely geometric quantity to the underlying many-body
physics, we compare the Wasserstein response with the electronic double
occupancy, a conventional measure of local charge correlations in the
Hubbard model. As shown in
Fig.~\ref{fig:wasserstein},
the strongest redistribution of coupling weight coincides with the region in
which double occupancy changes most rapidly. Although one quantity is defined
entirely from the encoded Hamiltonian and the other from the many-body ground
state, both identify the same interaction regime as the point of maximum
reorganization.

This agreement is significant. The Wasserstein distance is constructed solely
from the geometry of the encoded Hamiltonian and does not require solving the
many-body eigenvalue problem. Nevertheless, it reproduces the characteristic
interaction scale associated with the evolution of electronic correlations.
The XBK representation therefore provides an independent geometric
description of interaction-driven many-body physics that complements the
spectral analysis of the Bravyi--Kitaev hypergraph.

\section{Universality of the Geometric Framework}
\label{sec:universality}

The results presented thus far have focused on the Hubbard model, where the
competition between kinetic energy and Coulomb repulsion provides a natural
setting for studying interaction-driven geometric reorganization. An
important question, however, is whether the geometric framework developed in
the preceding sections reflects properties unique to the Hubbard model or
captures a more general organization of encoded quantum Hamiltonians.

To address this question, we apply exactly the same construction to four
additional many-body models spanning distinct classes of correlated quantum
systems: the one-dimensional Hubbard model, the spinless $t$--$V$ model,
the single-impurity Anderson model (SIAM), and the Kitaev chain. These
models encompass interaction-driven metal-insulator crossovers, impurity
physics, and topological phase transitions, thereby providing a stringent
test of the generality of the proposed framework.

For every Hamiltonian, we construct the corresponding BK hypergraph and XBK
coupling distribution using precisely the same procedure described in the
previous sections. No model-specific modifications are introduced. The same
spectral observables and the same Wasserstein geometry are therefore applied
uniformly across all systems.

\begin{figure*}[!tp]
\centering
\includegraphics[width=\linewidth]{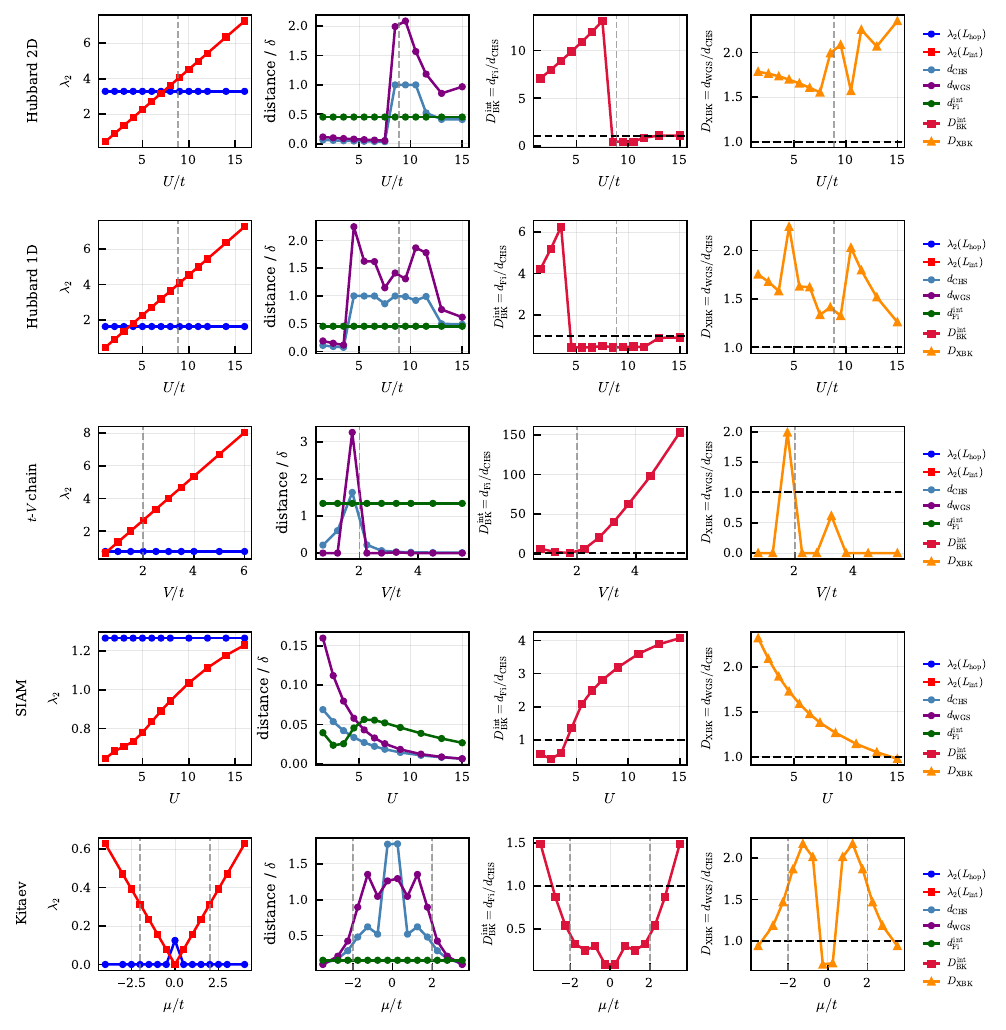}
\caption{\justifying 
\textbf{Universality of the geometric framework across correlated quantum
models.}
Rows correspond to the two-dimensional Hubbard model, one-dimensional
Hubbard model, spinless $t$--$V$ model, single-impurity Anderson model, and
Kitaev chain. For each Hamiltonian the left panels show the evolution of the
BK spectral observables, the middle panels compare the quantum-state and
coupling-space distances, and the right panels present the normalized
geometric response. Despite the different microscopic origins of their phase
transitions or crossovers, all models exhibit enhanced geometric
reorganization in the physically relevant parameter regime.
}
\label{fig:multimodel}
\end{figure*}
Figure~\ref{fig:multimodel} demonstrates that the geometric framework is not
restricted to the Hubbard Hamiltonian. Although these models differ
substantially in their microscopic interactions and physical mechanisms,
their encoded Hamiltonians exhibit a remarkably similar geometric response.

For the two-dimensional and one-dimensional Hubbard models, the spectral and
transport observables identify the interaction regime associated with the
onset of strong electronic correlations. The spinless $t$--$V$ model
displays an analogous enhancement near its interaction-driven transition,
despite the absence of spin degrees of freedom. In the SIAM, where no true
quantum phase transition occurs, the geometric observables develop a broad
maximum reflecting the continuous reorganization associated with Kondo
screening. Finally, for the Kitaev chain, both the spectral and
coupling-space measures respond sharply in the vicinity of the topological
phase transition.

These examples demonstrate that the proposed framework is sensitive not to a
particular microscopic Hamiltonian but to the structural reorganization of
the encoded interaction network. Whether the underlying physics is driven by
strong correlations, impurity effects, or topological order, the BK and XBK
geometries consistently detect the parameter regime in which the encoded
Hamiltonian undergoes its most significant reorganization.

The consistency observed across these diverse quantum many-body models
suggests that hypergraph geometry captures a general property of
fermion-to-qubit encodings rather than a feature specific to the Hubbard
model. The Bravyi--Kitaev and Xia--Bian--Kais representations therefore
provide complementary geometric languages that remain applicable across a
broad class of encoded quantum Hamiltonians. This universality considerably
extends the scope of the present framework beyond its original motivation in
strongly correlated electron systems.

\section{Conclusion}
\label{sec:conclusion}

We have developed a geometric framework for analyzing exact
fermion-to-qubit encodings based on weighted hypergraphs and coupling-space
representations. Rather than viewing an encoding solely as an intermediate
computational step for quantum simulation, we have shown that it defines a
well-structured geometric object whose organization reflects the underlying
many-body Hamiltonian. This perspective shifts the emphasis from the
wavefunction to the encoded operator itself and demonstrates that physically
meaningful information can be extracted directly from the geometry of the
encoding.

Within this framework we identified two complementary geometric
descriptions. The Bravyi--Kitaev representation naturally gives rise to a
hypergraph whose connectivity is characterized through spectral graph
theory, while the Xia--Bian--Kais representation provides a complementary
geometry in coupling space described by probability distributions and
optimal transport. Although these two representations probe different
aspects of the encoded Hamiltonian, they consistently describe the same
interaction-driven reorganization from complementary viewpoints.

The geometric framework reveals several unexpected structural properties of
fermion-to-qubit encodings. We introduced a dimensionless geometric
observable that compares the algebraic connectivities of the kinetic and
interaction hypergraphs and showed that it follows an exact analytical
dependence on the interaction strength. We further demonstrated that the
Bravyi--Kitaev encoding separates into two distinct geometric universality
classes and uncovered an exact internal spectral partition in which one
quarter of the interaction modes belong to a tree-dominated sector inherited
from the binary-tree architecture of the encoding. Finally, the
Xia--Bian--Kais representation showed that the same interaction-driven
reorganization can be quantified through the evolution of coupling
distributions using optimal transport. The consistency of these geometric
descriptions across the Hubbard, spinless $t$--$V$, Anderson impurity, and
Kitaev models suggests that the framework captures general features of
encoded quantum Hamiltonians rather than properties specific to a particular
many-body system.

More broadly, this work suggests that fermion-to-qubit mappings possess a
dual role. Beyond enabling quantum computation, they provide geometric
representations of many-body Hamiltonians whose structural organization
contains physically meaningful information. Hypergraph connectivity and
coupling-space transport therefore emerge as complementary geometric
languages for investigating encoded quantum systems. We expect that this
viewpoint can be extended to other fermion-to-qubit transformations~\cite{Ver05,Bal05,Set17,Set18,Hag23},
quantum-chemical Hamiltonians~\cite{Mni20,Mot20}, variational quantum algorithms~\cite{Per14,Mcc15,Mcc17abs}, and quantum
optimization problems~\cite{Luc14,Farhi2014,Hau20}, where geometric descriptors may provide new insights
into the structure of encoded many-body physics.

\section{Data Availability}
\label{app:data_availability}

The data and code that support the findings of this study are available from the corresponding author upon reasonable request.

\begin{acknowledgments}
The authors acknowledge support from the National Quantum Mission (NQM), an initiative of the Department of Science and Technology (DST), Government of India under the Project titled \emph{Q-LAT Anneal A - General-Purpose QUBO Compiler for Quantum Many-Body Physics and
Hybrid Optimisation}.
L.N. and N.R. were supported by the DAE Research Visitor Fellowship at The Institute of Mathematical Sciences (IMSc), Chennai. We also thank IMSc for providing HPC resources.
\end{acknowledgments}

\bibliographystyle{apsrev4-2}   
\bibliography{references}

@article{Nag26, author={Nagpal, L. and Kumar, A. and Hassan, S.R.}, title= {Mapping of fermionic lattice models for Ising solvers}, journal={Scientific Reports}, volume={16}, pages={16249}, year={2026}, doi = {10.1038/s41598-026-44886-7}}

@article{Sch11, author={Schollwock, U.}, title={The density-matrix renormalization group in the age
of matrix product states}, journal={Annals of Physics}, volume={326}, pages={96–192}, year={2011},
doi={10.1016/j.aop.2010.09.012}}

@article{Oru14, author={Orus, R.}, title={A Practical Introduction to
Tensor Networks: Matrix Product States and Projected Entangled Pair States}, journal={Annals of
Physics}, year={2014}, doi={10.1016/j.aop.2014.06.013}}

@article{Tro05, author={Troyer, M. and Wiese, U.-
J.}, title={Computational complexity and fundamental limitations to fermionic quantum Monte Carlo
simulations}, journal={Physical Review Letters}, volume={94}, pages={170201}, year={2005}, doi={10.1103/
PhysRevLett.94.170201}}

@article{Whi92, author={White, S. R.}, title={Density matrix formulation for
quantum renormalization groups}, journal={Physical Review Letters}, volume={69}, pages={2863–2866},
year={1992}, doi={10.1103/PhysRevLett.69.2863}}

@article{Pre18, author={Preskill, J.}, title={Quantum
Computing in the NISQ era and beyond}, journal={Quantum}, volume={2}, pages={79}, year={2018},
doi={10.22331/q-2018-08-06-79}}

@article{Luc14, author={Lucas, A.}, title={Ising formulations of many
NP problems}, journal={Frontiers in Physics}, volume={2}, pages={5}, year={2014}, doi={10.3389/
fphy.2014.00005}}

@misc{Boo20, author={Boothby, K. and Bunyk, P. I. and Raymond, J. and Roy, A.},
title={Next-Generation Topology of D-Wave Quantum Processors}, year={2020}, note={arXiv preprint}}

@article{Jor28, author={Jordan, P. and Wigner, E.}, title={Uber das Paulische Aquivalenzverbot}, journal={Zeitschrift
fur Physik}, volume={47}, pages={631–651}, year={1928}, doi={10.1007/BF01331938}}

@article{Bra00,
author={Bravyi, S. and Kitaev, A.}, title={Fermionic quantum computation}, journal={Annals
of Physics}, volume={298}, pages={210–226}, year={2002}, doi={10.1006/aphy.2002.6254}}

@article{See12,
author={Seeley, J. T. and Richard, M. J. and Love, P. J.}, title={The Bravyi–Kitaev transformation for
quantum computation of electronic structure}, journal={The Journal of Chemical Physics}, volume={137},
pages={224109}, year={2012}, doi={10.1063/1.4768229}}

@article{Llo96, author={Lloyd, S.}, title={Universal
Quantum Simulators}, journal={Science}, volume={273}, pages={1073–1078}, year={1996}, doi={10.1126/
science.273.5278.1073}}

@article{Per14, author={Peruzzo, A. and McClean, J. R. and Shadbolt, P. and Yung,
M.-H. and Zhou, X.-Q. and Love, P. J. and Aspuru-Guzik, A. and OBrien, J. L.}, title={A variational
eigenvalue solver on a photonic quantum processor}, journal={Nature Communications}, volume={5},
pages={4213}, year={2014}, doi={10.1038/ncomms5213}}

@article{Hav17, author={Havlicek, V. and Troyer,
M. and Whitfield, J. D.}, title={Operator locality in the quantum simulation of fermionic models}, journal={Physical
Review A}, volume={95}, pages={032332}, year={2017}, doi={10.1103/PhysRevA.95.032332}}

@article{Ste19, author={Steudtner, M. and Wehner, S.}, title={Quantum codes for quantum simulation
of fermions on a square lattice of qubits}, journal={Physical Review A}, volume={99}, pages={022308},
year={2019}, doi={10.1103/PhysRevA.99.022308}}

@article{Xia21, author={Xia, R. and Bian, T. and Kais, S.},
title={Electronic structure calculations and the Ising Hamiltonian / XBK method}, journal={The Journal of
Chemical Physics}, volume={154}, pages={014106}, year={2021}}

@article{Xia17, author={Xia, R. and Bian,
T. and Kais, S.}, title={Electronic Structure Calculations and the Ising Hamiltonian}, journal={The Journal
of Physical Chemistry B}, year={2017}, doi={10.1021/acs.jpcb.7b10371}}

@misc{Bra17, author={Bravyi,S. and Gambetta, J. M. and Mezzacapo, A. and Temme, K.}, title={Tapering off qubits to simulate
fermionic Hamiltonians}, year={2017}, eprint={1701.08213}, archivePrefix={arXiv}, primaryClass={quant-
ph}}

@article{Ros75, author={Rosenberg, I. G.}, title={Reduction of bivalent maximization to the quadratic
case}, journal={Cahiers du Centre d’Etudes de Recherche Operationnelle}, volume={17}, pages={71–
74}, year={1975}}

@article{Ish11, author={Ishikawa, H.}, title={Transformation of General Binary MRF
Minimization to the First-Order Case}, journal={IEEE Transactions on Pattern Analysis and Machine
Intelligence}, volume={33}, pages={1234–1249}, year={2011}}

@article{Din67, author={Dinkelbach, W.},
title={On Nonlinear Fractional Programming}, journal={Management Science}, volume={13}, pages={492–
498}, year={1967}}

@article{Cho11a, author={Choi, V.}, title={Reducing the QUBO problem to the
Ising model}, journal={Quantum Information Processing}, volume={10}, pages={343–353}, year={2011}}

@article{Cho11b, author={Choi, V.}, title={Minor-embedding in adiabatic quantum computation: I. The
parameter setting problem}, journal={Quantum Information Processing}, volume={10}, pages={193–209},
year={2008}, doi={10.1007/s11128-008-0082-9}}

@article{Bor02, author={Boros, E. and Hammer, P. L.}, title={Pseudo-Boolean
optimization}, journal={Discrete Applied Mathematics}, volume={123}, pages={155–
225}, year={2002}, doi={10.1016/S0166-218X(01)00341-9}}

@misc{Glo18, author={Glover, F. and Kochenberger,
G. A. and Du, Y.}, title={A Tutorial on Formulating and Using QUBO Models}, year={2018},
eprint={1811.11538}, archivePrefix={arXiv}, primaryClass={quant-ph}}

@article{Hau20, author={Hauke, P.
and Katzgraber, H. G. and Lechner, W. and Nishimori, H. and Oliver, W. D.}, title={Perspectives of quantum
annealing: methods and implementations}, journal={Reports on Progress in Physics}, volume={83},
pages={054401}, year={2020}, doi={10.1088/1361-6633/ab85b8}}

@phdthesis{Got97, author={Gottesman,
D.}, title={Stabilizer Codes and Quantum Error Correction}, school={California Institute of Technology},
year={1997}, doi={10.7907/RZR7-DT72}}

@article{Aar05, author={Aaronson, S. and Gottesman,
D.}, title={Improved simulation of stabilizer circuits}, journal={Proceedings of the Royal Society A}, volume={461},
pages={3473–3482}, year={2005}, doi={10.1098/rspa.2005.1546}}

@book{NielsenChuang2010,
author={Nielsen, M. A. and Chuang, I. L.}, title={Quantum Computation and Quantum Information},
edition={10}, publisher={Cambridge University Press}, year={2010}}

@article{Kad98, author={Kadowaki, T.
and Nishimori, H.}, title={Quantum annealing in the transverse Ising model}, journal={Physical Review
E}, volume={58}, pages={5355–5363}, year={1998}, doi={10.1103/PhysRevE.58.5355}}

@article{Huk96, author={Hukushima,
K. and Nemoto, K.}, title={Exchange Monte Carlo Method and Application to Spin
Glass Simulations}, journal={Journal of the Physical Society of Japan}, volume={65}, pages={1604–1608},
year={1996}, doi={10.1143/JPSJ.65.1604}}

@article{Rnn14, author={Ronnnow, T. F. and Wang, Z. and Job,
J. and Boixo, S. and Isakov, S. V. and Wecker, D. and Svore, K. M. and Troyer, M.}, title={Defining and detecting
quantum speedup}, journal={Science}, volume={345}, pages={420–424}, year={2014}, doi={10.1126/
science.1252319}}

@misc{Farhi2000, author={Farhi, E. and Goldstone, J. and Gutmann, S. and Sipser,
M.}, title={Quantum Computation by Adiabatic Evolution}, year={2000}, eprint={quant-ph/0001106},
archivePrefix={arXiv}, primaryClass={quant-ph}}

@article{Farhi2001, author={Farhi, E. and Goldstone, J.
and Gutmann, S. and Lapan, J. M. and Lundgren, A. and Preda, D.}, title={A Quantum Adiabatic Evolution
Algorithm Applied to Random Instances of an NP-Complete Problem}, journal={Science}, year={2001},
doi={10.1126/science.1057726}}

@misc{Farhi2014, author={Farhi, E. and Goldstone, J. and Gutmann, S.},
title={A Quantum Approximate Optimization Algorithm}, year={2014}, eprint={1411.4028}, archivePrefix={arXiv},
primaryClass={quant-ph}}

@article{Whi93, author={White, S. R.}, title={Density-matrix algorithms
for quantum renormalization groups}, journal={Physical Review B}, volume={48}, pages={10345–10356}, year={1993}, doi={10.1103/PhysRevB.48.10345}}

@inproceedings{Mcg13, author={McGeoch, C.
and Wang, C.}, title={Experimental Evaluation of an Adiabatic Quantum System for Combinatorial
Optimization}, booktitle={Proceedings of the ACM International Conference on Computing Frontiers},
year={2013}, doi={10.1145/2482767.2482797}}

@article{Ven15, author={Venturelli, D. and Mandra, S. and
Knysh, S. and OGorman, B. and Biswas, R. and Smelyanskiy, V.}, title={Quantum Optimization of
Fully-Connected Spin Glasses}, journal={Physical Review X}, volume={5}, pages={031040}, year={2015},
doi={10.1103/PhysRevX.5.031040}}

@article{Kin18, author={King, A. D. and Carrasquilla, J. and Raymond,
J. and Ozfidan, I. and Andriyash, E. and Berkley, A. J. and Reis, M. and Lanting, T. and Harris, R.
and Altman, E. and Amin, M. H.}, title={Observation of topological phenomena in a programmable
lattice of 1,800 qubits}, journal={Nature}, volume={560}, pages={456–460}, year={2018}, doi={10.1038/
s41586-018-0410-x}}

@article{Alb18, author={Albash, T. and Lidar, D. A.}, title={Adiabatic quantum computation},
journal={Reviews of Modern Physics}, volume={90}, pages={015002}, year={2018}, doi={10.1103/
RevModPhys.90.015002}}

@article{Asp05, author={Aspuru-Guzik, A. and Dutoi, A. D. and Love, P. J. and
Head-Gordon, M.}, title={Simulated quantum computation of molecular energies}, journal={Science}, volume={309},
pages={1704–1707}, year={2005}, doi={10.1126/science.1113479}}

@article{Cao19, author={Cao,
Y. and Romero, J. and Olson, J. P. and Degroote, M. and Johnson, P. D. and Kieferova, M. and Kivlichan, I.
D. and Menke, T. and Peropadre, B. and Sawaya, N. P. D. and Sim, S. and Veis, L. and Aspuru-Guzik, A.},
title={Quantum Chemistry in the Age of Quantum Computing}, journal={Chemical Reviews}, year={2019},
doi={10.1021/acs.chemrev.8b00803}}

@article{Fey99, author={Feynman, R. P.}, title={Simulating physics
with computers}, journal={International Journal of Theoretical Physics}, volume={21}, pages={467–
488}, year={1999}, doi={10.1007/BF02650179}}

@article{Fin94, author={Finnila, A. B. and Gomez, M. A.
and Sebenik, C. and Stenson, C. and Doll, J. D.}, title={Quantum annealing: A new method for
minimizing multidimensional functions}, journal={Chemical Physics Letters}, volume={219}, pages={343–
348}, year={1994}, doi={10.1016/0009-2614(94)00117-0}}

@misc{Hag23, author={Hagge, T. and Wiebe,
N.}, title={Error mitigation via error detection using Generalized Superfast Encodings}, year={2023}}

@misc{Mca18, author={McArdle, S. and Endo, S.
and Aspuru-Guzik, A. and Benjamin, S. C. and Yuan, X.}, title={Quantum computational chemistry},
year={2018}}

@article{Mcc15,
author={McClean, J. R. and Romero, J. and Babbush, R. and Aspuru-Guzik, A.}, title={The theory of
variational hybrid quantum-classical algorithms}, journal={New Journal of Physics}, volume={18}, number={2},
pages={023023}, year={2016}, doi={10.1088/1367-2630/18/2/023023}}

@inproceedings{Mcc17abs,
author={McClean, J. R. and Rubin, N. C. and Kivlichan, I. D. and Sung, K. J. and Steiger, D. S. and
Bonet-Monroig, X. and Cao, Y. and Dai, C. and Fried, E. S. and Gidney, C. and Gokhale, P. and Haner,
T. and Hardikar, T. S. and Havlicek, V. and Higgott, O. and Huang, C. and Izaac, J. and Jiang, Z. and
Liu, X. and McArdle, S. and Neeley, M. and OBrien, T. E. and OGorman, B. and Ozfidan, I. and Radin,
M. D. and Romero, J. and Sawaya, N. P. D. and Setia, K. and Sim, S. and Steudtner, M. and Sun, W.
and Wang, D. and Zhang, F. and Babbush, R.}, title={OpenFermion: The Electronic Structure Package for
Quantum Computers}, booktitle={Bulletin of the American Physical Society}, year={2017}}

@misc{Mni20,
author={Mniszewski, S. M. and Dub, P. A. and Tretiak, S. and Anisimov, P. M. and Zhang, Y. and
Negre, C. F. A.}, title={Downfolding the Molecular Hamiltonian Matrix using Quantum Community
Detection}, year={2020}, note={arXiv preprint}}

@article{Mot20, author={Motta, M. and Sun, C. and Tan,
A. T. K. and ORourke, M. J. and Ye, E. and Minnich, A. J. and Brandao, F. G. S. and Chan, G. K.-L.}, title={Determining eigenstates and thermal states on a quantum computer using quantum imaginary
time evolution}, journal={Nature Physics}, year={2020}, doi={10.1038/s41567-019-0704-4}}

@article{Rol01,
author={Roland, J. and Cerf, N. J.}, title={Quantum search by local adiabatic evolution}, journal={Physical
Review A}, volume={65}, pages={042308}, year={2002}, doi={10.1103/PhysRevA.65.042308}}

@article{Set17,
author={Setia, K. and Whitfield, J. D.}, title={Bravyi–Kitaev Superfast simulation of electronic structure
on a quantum computer}, journal={The Journal of Chemical Physics}, pages={164104}, year={2018},
doi={10.1063/1.5019371}}

@article{Set18, author={Setia, K. and Bravyi, S. and Mezzacapo, A. and Whitfield,
J. D.}, title={Superfast encodings for fermionic quantum simulation}, journal={Physical Review Research},
volume={1}, pages={033033}, year={2019}, doi={10.1103/PhysRevResearch.1.033033}}

@article{Tep19, author={Teplukhin,
A. and Kendrick, B. K. and Babikov, D.}, title={Calculation of Molecular Vibrational
Spectra on a Quantum Annealer}, journal={Journal of Chemical Theory and Computation}, year={2019},
doi={10.1021/acs.jctc.9b00402}}

@article{Tep20, author={Teplukhin, A. and Kendrick, B. K. and Tretiak,
S. and Dub, P. A.}, title={Electronic structure with direct diagonalization on a D-Wave quantum
annealer}, journal={Scientific Reports}, year={2020}, doi={10.1038/s41598-020-77315-4}}

@article{Tep21,
author={Teplukhin, A. and Kendrick, B. K. and Mniszewski, S. M. and Zhang, Y. and Kumar, A. and Negre,
C. F. A. and Anisimov, P. M. and Tretiak, S. and Dub, P. A.}, title={Computing molecular excited states on
a D-Wave quantum annealer}, journal={Scientific Reports}, year={2021}, doi={10.1038/s41598-021-98331-
y}}

@article{Tep22, author={Teplukhin, A. and Kendrick, B. K. and Mniszewski, S. M. and Tretiak,
S. and Dub, P. A.}, title={Sampling electronic structure quadratic unconstrained binary optimization
problems (QUBOs) with Ocean and Mukai solvers}, journal={PLOS ONE}, year={2022}, doi={10.1371/
journal.pone.0263849}}

@inproceedings{Ush17, author={Ushijima-Mwesigwa, H. and Negre, C. F. A.
and Mniszewski, S. M.}, title={Graph Partitioning using Quantum Annealing on the D-Wave System},
booktitle={Proceedings of the ACM International Conference on Computing Frontiers}, year={2017},
doi={10.1145/3149526.3149531}}

@article{Ver05, author={Verstraete, F. and Cirac, J. I.}, title={Mapping
local Hamiltonians of fermions to local Hamiltonians of spins}, journal={Journal of Statistical Mechanics:
Theory and Experiment}, pages={P09012}, year={2005}, doi={10.1088/1742-5468/2005/09/P09012}}

@article{Bal05,
author={Ball, R. C.}, title={Fermions without Fermion Fields}, journal={Physical Review Letters},
volume={95}, pages={176407}, year={2005}, doi={10.1103/PhysRevLett.95.176407}}

\clearpage
\appendix
\appendix
\section{Computational Methods}
\label{app:methods}

All results are obtained by classical simulation. This appendix specifies
the numerical methods and system sizes used to generate every figure and
table. Fermion-to-qubit mappings use a standard open-source transformation
package; linear algebra uses NumPy/SciPy; optimal-transport linear programs
are solved exactly with the HiGHS solver. Unless stated otherwise $t=1$ and
all lattices are at half filling with periodic boundary conditions (PBC).

\subsection{Model and encoding}
\label{app:model_bk}

The 2D Hubbard model on an $L_x\times L_y$ lattice is
\begin{equation}
  H = -t\!\sum_{\langle i,j\rangle,\sigma}\!(c^{\dagger}_{i\sigma}c_{j\sigma}+\mathrm{h.c.})
      + U\sum_{i} n_{i\uparrow}n_{i\downarrow} - \mu\sum_{i,\sigma} n_{i\sigma},
\end{equation}
with half filling ($N_e=L_xL_y$, $\mu=0$). Comparison models (1D Hubbard,
spinless $t$--$V$ chain, SIAM, Kitaev chain) use analogous conventions; the
Kitaev chain is number non-conserving and is treated without a
particle-number penalty or tapering.

Each Hamiltonian is mapped to a qubit operator via the Bravyi--Kitaev
transform. Every Pauli term is assigned to $H_{\rm hop}$ if it contains any
$X$ or $Y$ factor, and to $H_{\rm int}$ if it is pure-$Z$; this split is
exact. Where a fixed particle sector is required, a quadratic penalty
$H\to H + w(N_e-\sum_i n_i)^2$ ($w=5$) is added \emph{before} encoding to
select the half-filled sector; all reported observables are then evaluated
on the penalty-free Hamiltonian restricted to that sector.

\subsection{Tapering}
\label{app:taper}

$\mathbb{Z}_2$ qubit tapering~\cite{Bra17} is performed by writing each
Pauli string as a symplectic row over $\mathrm{GF}(2)$, taking the kernel of
the resulting parity-check matrix to obtain independent symmetry generators,
and projecting the Hamiltonian onto the sector (over all $\pm1$ generator
sign choices) of lowest ground energy. Tapering is like-for-like only among
sizes sharing the same generator structure; the tapered family used
throughout is $2\times2,\,2\times3,\,2\times4$ ($n_q\le14$ post-taper), which
is what makes exact ground states tractable. A fourth tapered geometry,
$3\times3$, was also computed but is excluded from the $\rOT$ crossing
analysis: odd$\times$odd PBC introduces lattice frustration, and $\rOT$ does
not cross 1 within $U/t\le20$ for this size ($\Ustar$ undefined over the
studied range).

\subsection{Hypergraph Laplacian and clique expansion}
\label{app:laplacian}

Each qubit is a vertex; each non-identity Pauli term of coefficient
$c_\alpha$ is a hyperedge $S_\alpha$ over its support. The weighted Laplacian
is
\begin{equation}
  L_{ij} = -\!\!\sum_{\alpha:\,i,j\in S_\alpha}\!\!
           \frac{|c_\alpha|\,w_\alpha(i)\,w_\alpha(j)}{|S_\alpha|-1}
  \ (i\neq j), \qquad
  L_{ii} = \sum_{j\neq i} |L_{ij}|,
  \label{eq:app_laplacian}
\end{equation}
the standard graph-Laplacian sign convention ($L=D-A$), symmetric positive
semi-definite with zero row sums. The Pauli weight is $w_\alpha(i)=1$ for a
$Z$ leg and $\tfrac12$ for an $X$ or $Y$ leg. A $k$-body term is clique-expanded
onto its $\binom{k}{2}$ pairs, coupling divided by $(k-1)$; shared pairs
accumulate contributions. No auxiliary qubits are introduced.

The Fiedler value $\lambda_2$ is the smallest eigenvalue above a $10^{-10}$
null-space threshold. The structural ratio is $\rOT(U)=\lambda_2(\Lhop)/\lambda_2(\Lint(U))$;
since $\lambda_2(\Lint(U))=\text{slope}\cdot U$ by construction, the crossing
is $\Ustar=\lambda_2(\Lhop)/\text{slope}$. The spectral-family analysis
applies this to every eigenvalue index $k$, with slopes extracted at a fixed
reference $U_0/t=1$.

For the coupling-space transport diagnostic, each hyperedge additionally
defines a ground-metric edge length $\ell^{(\alpha)}_{ij}=|S_\alpha|/|c_\alpha|$
(shorter for stronger couplings); shared pairs keep the minimum length. The
resulting graph distance $d(i,j)$ is the all-pairs shortest path (Dijkstra),
with disconnected pairs at $d=\infty$ and a $10^{-12}$ regularizer guarding
the coefficient division.

\subsection{Eigenvalue solves and system sizes}
\label{app:eig}

All eigensolves reported in the paper are dense (\texttt{eigvalsh}),
affordable since $n_q$ never exceeds a few hundred in any submitted figure.
Table~\ref{tab:app_sizes} lists exactly the sizes used.

\begin{table}[t]
\centering
\caption{\justifying System sizes used in each figure. $n_q$ is the BK qubit count
  (Laplacian dimension), untapered unless noted.}
\label{tab:app_sizes}
\begin{ruledtabular}
\begin{tabular}{lll}
Dataset & Sizes ($L_x\times L_y$) & Solver \\
\hline
Structural $\rOT$ sweep & $2\times2,2\times3,2\times4,3\times3$ & dense \\
Spectral family & $4\times4,4\times6,8\times8,$ & dense \\
  & $10\times10,12\times12,16\times16$ & (full spectrum)\\
Tapered strips / ED & $2\times2,2\times3,2\times4$ & sparse \texttt{eigsh}\\
\end{tabular}
\end{ruledtabular}
\end{table}

The $U$ grid is non-uniform, denser near the crossover; $\Ustar$ is obtained
by linear interpolation between the two bracketing grid points ($\pm0.5$
resolution near the crossing), which propagates into the FSS uncertainty
below.

\subsection{Exact diagonalization and Wasserstein-1 distance}
\label{app:ed_wass}

Ground states for the Wasserstein and double-occupancy diagnostics come from
exact diagonalization of the tapered, penalty-free Hamiltonian ($n_q\le14$),
using the eigenvector of least eigenvalue. Probabilities are truncated to
support $p(x)>10^{-10}$ (renormalized). Double occupancy is
$D(U)=\tfrac1N\partial E_0/\partial U$ via central finite difference.

Two distinct $W_1$ computations appear, both solved as exact LPs (no
entropic regularization). Between ground states, the ground metric is
Hamming distance on the Boolean hypercube,
$d_H(x,y)=\mathrm{popcount}(x\oplus y)$; the LP is solved directly over the
truncated support (top-$m{=}200$ by mass for larger strips). On the coupling
graph itself, the ground metric is instead the shortest-path distance
described in Sec.~\ref{app:laplacian}, with lazy random-walk measures
(laziness $\alpha=\tfrac12$) and curvature $\kappa(i,j)=1-W_1(\mu_i,\mu_j)/d(i,j)$.
A failed LP solve returns zero and is flagged.

\subsection{Finite-size scaling}
\label{app:fss}

The tapered crossings ($\Ustar/t = 7.26, 9.51, 9.89$ for $N=4,6,8$) are fit
to $\Ustar(L) = U_c + a/\sqrt{N_{\rm sites}}$. With only three points, this
is a one-degree-of-freedom linear regression, and we report the resulting
uncertainty honestly: $U_c/t = 8.87\pm2.18$ (1$\sigma$), with a wide
confidence interval reflecting the severity of a three-point extrapolation.
We do not claim precise quantitative agreement with quantum Monte Carlo
estimates of $U_c$ on the strength of this fit alone; the more robust
statement is the trend that crossings increase systematically with strip
width, together with the individual dimensionless ratios $\alpha=\Ustar/W$
landing near unity for each tapered strip. Untapered 2D lattices, which
cross $\rOT=1$ only at substantially higher $U/t$ on a different qubit set,
are excluded from the fit and shown only as comparison points. Additional
tapered sizes would be needed to meaningfully tighten this interval.

\subsection{Reproducibility}
\label{app:env}

Simulations ran in a fixed Python environment (NumPy, SciPy/HiGHS, a graph
library for shortest paths, a fermion-to-qubit transformation package) on
institutional HPC resources. All computation is deterministic except where
randomness enters (e.g., bootstrap resampling), for which a fixed seed is
used. Every figure is reproducible from its corresponding dataset via
deterministic post-processing.
\end{document}